\documentclass[prb,showpacs,twocolumn]{revtex4-1}

\usepackage{graphicx}
\usepackage{amsmath}

\usepackage{color}

\usepackage{nicefrac}

\begin{document}

\title{Universal quench dynamics of interacting quantum impurity systems}

\author{D.M.\ Kennes$^1$, V.\ Meden$^1$ and R.\ Vasseur$^{2,3}$}

\address{$^1$Institut f\"ur Theorie der Statistischen Physik, RWTH Aachen University 
and JARA---Fundamentals of Future Information Technology, 52056 Aachen, Germany\\
$^2$Department of Physics, University of California, Berkeley, Berkeley CA 94720, U.S.A.\\
$^3$Materials Science Division, Lawrence Berkeley National Laboratory, Berkeley CA 94720, U.S.A.}

\begin{abstract}
The {\em equilibrium} physics of quantum impurities frequently involves a 
universal crossover from weak to strong reservoir-impurity coupling, 
characterized by single-parameter scaling and an energy scale 
$T_K$ (Kondo temperature) that breaks scale invariance. For the 
\emph{non-interacting} resonant level model, 
the {\em non-equilibrium} time evolution of the Loschmidt echo after a local 
quantum quench was recently computed explicitly [R. Vasseur, K. Trinh, 
S. Haas, and H. Saleur, Phys.~Rev.~Lett.~{\bf 110}, 240601 (2013)]. 
It shows single-parameter scaling with variable $T_K t$.
Here, we scrutinize whether similar universal dynamics can be observed in various 
{\em interacting} quantum impurity systems. Using density matrix and functional 
renormalization group approaches, we analyze the time evolution 
resulting from abruptly 
coupling two non-interacting Fermi or interacting Luttinger liquid leads via a 
quantum dot or a direct link. We also consider the case of a single Luttinger 
liquid lead suddenly coupled to a quantum dot. We investigate whether the 
{\em field theory} predictions for the universal scaling as well as for the large time 
behavior successfully describe the time evolution of the Loschmidt echo and the 
entanglement entropy of {\em microscopic models.} 
Our study shows that for the considered local quench protocols the above quantum impurity 
models fall into a class of problems for which the non-equilibrium dynamics can 
largely be understood based on the knowledge of the corresponding equilibrium physics.    
\end{abstract}


\maketitle

\section{Introduction}

In condensed matter physics non-equilibrium problems have attracted increasing interest in recent 
years. Despite many advances, describing those problems accurately in the presence of strong 
correlations remains a formidable challenge even today. The physics is often involved  and
the number of available non-equilibrium many-body methods allowing for a controlled access beyond 
plain perturbation theory is limited. Furthermore, each of those has its own advantages and 
shortcomings. An alternative way to gain insights into the interplay of non-equilibrium and 
correlations is to identify non-equilibrium problems which can largely be understood in terms 
of their equilibrium physics; compared to non-equilibrium, the latter is frequently rather 
well understood. Generally, scrutinizing under which conditions such a reduction in complexity 
can be applied, is of importance to draw a more complete picture. 
Here we conduct such a study for the non-equilibrium physics of so called quantum quenches
in quantum impurity problems.  

As they provide a basic probe of the non-equilibrium dynamics of many-body quantum systems, 
quantum quenches have been of great interest recently. This development is motivated by the 
progress of experiments with ultra-cold atoms in a tunable potential.~\cite{Cold1} 
At time $t=0$, the system is suddenly brought far from equilibrium by abruptly changing a 
control parameter, and is subsequently left to evolve unitarily. {\em Global} quenches, for 
which the control parameter is quenched throughout the entire system, correspond to injecting 
an extensive amount of energy, thereby allowing one to address the intriguing issues of 
thermalization and non-equilibrium steady-states in closed many-body quantum systems.~\cite{Global1} 
{\em Local} quantum quenches, for which the sudden change is restricted in 
space,~\cite{CardyCalabrese2b, CardyCalabrese2,EE1b,Fagotti, Laeuchli, EchoCFT2,VM_QSH} 
allow one to investigate the energy propagation. One might expect that for large times only 
the low-energy excitations matter and the non-equilibrium dynamics shows universality, 
provided the system in equilibrium has this property. We here provide evidence that for several 
quantum impurity problems this is indeed the case. 
In this work, we study a particular class of local quantum quenches: at time $t=0$, 
two independent metallic reservoirs are either directly tunnel-coupled or coupled through
a single-level quantum dot and left to evolve unitarily following the Schr\"odinger equation. 
In addition, we investigate the quench dynamics after abruptly tunnel coupling a single lead to a quantum dot.
A quantum quench of this type has recently been realized experimentally exploiting the optical 
absorption of a semiconductor quantum dot.~\cite{KondoQuench1,KondoQuench2,HKXray}
In equilibrium, the considered impurity setups are characterized by a low-energy scale $T_K$, the Kondo 
temperature.

From a theory perspective, it was argued ~\cite{LetterRLM} based on general considerations 
and explicit computations for a non-interacting impurity problem, that the energy scale $T_K$ characterizes  
a crossover in the post-quench dynamics, with the long-time behavior being essentially controlled 
by the low-energy properties of the Hamiltonian after the quench.  For the considered type of local 
quench, one can expect that the same holds also in the presence of two-particle interactions. 
We focus on parameter regimes in which the sub-system coupling (weak link or dot) 
is a relevant perturbation in the renormalization group (RG) sense in equilibrium, 
and therefore drastically alters the low-energy properties of the system. 
The explicit non-interacting result of Ref.~\onlinecite{LetterRLM} and the expectation for  
interacting systems can thus be interpreted as follows: time essentially acts as the inverse 
of an energy scale, and the post-quench dynamics effectively follows the RG flow. At large 
times, the system `heals' itself so that the quantum impurity becomes strongly hybridized 
with the lead(s).

\begin{figure}[tb]
 \centering
 \includegraphics[width=0.48\textwidth]{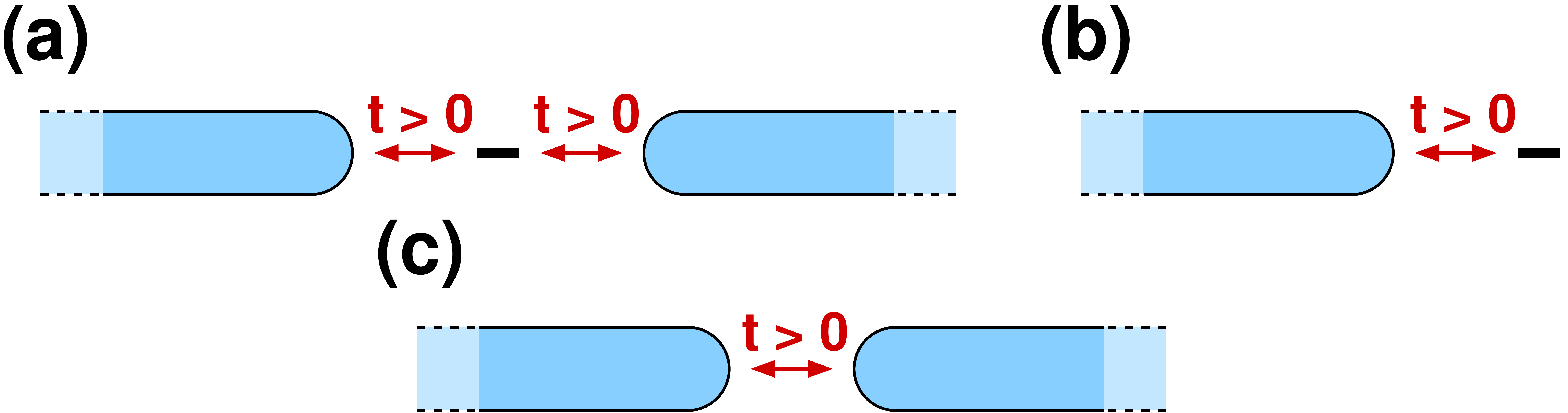}
\caption{(Color online) The non-equilibrium setups studied in this work: initially decoupled chains which are 
prepared in their respective ground states are coupled at $t=0$ via different contacts. We focus 
on two different of such geometries: (a) coupling the reservoirs indirectly by a single level 
quantum dot and (c) connecting  the two chains directly by a weak link, respectively. Furthermore, 
we consider the setup (b) in which a single lead is coupled to a single level quantum dot. 
The subsequent out-of-equilibrium dynamics is analyzed.}
  \label{fig:setup}
\end{figure}
Because of the energy scale $T_K$ that breaks conformal invariance, many of the analytical 
results on local quenches from conformal field theory~\cite{CardyCalabrese2} cannot be used. It is in 
general very difficult to provide closed analytical expressions for the time evolution of 
observables showing the full crossover, even for formally exactly solvable (`integrable') 
or non-interacting systems. However, if the above scenario holds, one can expect 
that the post-quench dynamics is universal in the sense that every observable only 
depends on $T_K t$ (in the absence of any other energy scale) and that the long-time regime 
$T_K t \gg 1$ can be described using boundary conformal field theory (BCFT). 
In this manuscript, we investigate numerically whether this is indeed the case for various interacting 
quantum impurity setups, using a combination of density matrix renormalization group 
(DMRG)~\cite{White, Vidal04, Schollwoeck11}  and functional renormalization group 
(FRG)~\cite{Metzner12,Kennes12,Kennes12a,Kennes13} methods. 

We focus on two key observables: the entanglement entropy between sub-systems, whose long-time 
behavior reduces to the CFT prediction of Refs.~\onlinecite{CardyCalabrese2} 
(see Ref.~\onlinecite{CalabreseWeakLink} for a similar observation), and the time dependent fidelity 
or Loschmidt echo,  whose long-time behavior carries signatures of the effective boundary 
condition felt by the lead(s). The latter is induced by the screened impurity. The Loschmidt 
echo is related to the equilibrium  Anderson orthogonality catastrophe~\cite{Anderson} 
and is of particular interest since it can be linked to the Fourier transform of the absorption 
spectrum measured in a recent experiment.~\cite{KondoQuench2} In all the examples considered, 
we find that the field theory predictions describe reasonably well the dynamics of the microscopic 
models under scrutiny.

The remainder of this paper is organized as follows. In Sect.~2 we introduce the different models 
and observables studied, along with the analytical and numerical methods used in our analysis. Section 3 
contains a detailed discussion of the so-called interacting resonant level model (IRLM), the arguably 
simplest interacting 
impurity model with spinless non-interacting Fermi liquid leads. In Sect.~4 we generalize this impurity 
setup to the richer case of interacting Luttinger liquid leads, with various impurity configurations 
(one or two leads tunnel-coupled to a quantum dot, two leads 
connected through a point contact weak link). Finally, Sect.~5 provides a summary of the main results as 
well as a discussion of perspectives for future work. 

\section{Models and Methods}

\subsection{Models}

We mainly consider setups in which two decoupled fermionic reservoirs prepared in their respective 
ground states 
are coupled to each other at time $t=0$. Thus the combined system is abruptly brought out of 
equilibrium at this time. This protocol constitutes a specific local quantum quench. We analyze 
different microscopic lattice models corresponding to various settings, with the 
reservoirs being either (single-channel) non-interacting Fermi liquids or one-dimensional (1d)
interacting Luttinger liquids, and the junction being either a single link or a  quantum dot. 
These setups are depicted schematically in Figs.~\ref{fig:setup} (a) and (c). Furthermore, we 
investigate the case of a single Luttinger liquid reservoir being coupled abruptly to a 
single level quantum impurity as shown in Fig.~\ref{fig:setup} (b).
  
\emph{IRLM} --- First, we investigate a microscopic realization of the IRLM 
(see e.g.~Refs.~\onlinecite{Schlottmann,Doyon,IRLM2,Karrasch10,Andergassen} as well as references 
therein), which describes a single localized charge level (quantum dot) of energy $\epsilon$ 
coupled to a left (index $\alpha={\rm L}$) and a right (index $\alpha={\rm R}$) non-interacting 
1d reservoir; see Fig.~\ref{fig:setup} (a). In our non-equilibrium setup the 
tunneling couplings  $\gamma_\alpha'$ between the quantum dot level and the reservoirs are abruptly 
turned on at time $t=0$. We consider a situation in which the two-particle interactions $U_\alpha$ 
between dot and lead fermions are active for all times. The microscopic Hamiltonian reads
\begin{align}
H=& H_{\rm{dot} }+\sum \limits_{\alpha={\rm L},{\rm R}}\left[ H_{\rm{coup},\alpha }(t)+H_{\rm{res},\alpha }\right],\\
H_{\rm{dot}}=& \epsilon \hat n_0,\\
H_{\rm{coup},\alpha}(t)=&\Theta(t) \gamma_\alpha^\prime \left(c_{1,\alpha}^\dagger 
c_0+c_0^\dagger c_{1,\alpha} \right) \notag
\\+&U_\alpha \left(\hat n_0- \frac12 \right) 
\left(\hat n_{1,\alpha} - \frac12 \right),\\
H_{\rm{res},\alpha}=& -\gamma_\alpha \sum \limits_{j=1}^{L-1} 
\left(c_{j+1,\alpha}^\dagger c_{j,\alpha}+c_{j,\alpha}^\dagger c_{j+1,\alpha}\right).
\end{align}
We use standard second quantization notation by introducing the fermionic annihilation (creation) 
operator $c_j$ ($c_j^\dagger$) of a spinless particle at site $j$. Furthermore, $\hat  n_{0}=
c_{0}^\dagger c_{0}$ and $\hat  n_{1,\alpha }=c_{1,\alpha }^\dagger c_{1,\alpha }$ denote the occupancy 
operators of the dot and the first site of reservoir $\alpha$, respectively. We have chosen a 
tight-binding description of the reservoirs of length $L$ with hopping $\gamma_\alpha>0$ and 
open boundary conditions. The total number of sites in such a setup is $2L+1$.

\emph{Interacting reservoirs} --- We also consider models with interacting reservoirs 
(rather than non-interacting Fermi liquids as in the IRLM). This is achieved by including 
nearest-neighbor interactions of strength $J_\alpha \Delta_\alpha$ in the 1d leads. 
The Hamiltonian is given by 
\begin{equation}
H=\sum\limits_{\alpha={\rm L,R}} H_{\rm{chain},\alpha}+H_{\rm{coup}}(t)
\label{eqXXZcase}
\end{equation}
with the chain part (nearest-neighbor hopping $J_\alpha/2>0$)
\begin{align}
H_{\rm{chain},\alpha}=&\sum\limits_{j=1}^{L-1} J_\alpha\left[\frac{1}{2}c_{j+1,\alpha}^\dagger c_{j,\alpha}+\frac{1}{2}c_{j,\alpha}^\dagger c_{j+1,\alpha} \right. \notag \\ &+ \left.  \Delta_\alpha \left(\hat n_{j,\alpha}- \frac12 \right) \left(\hat n_{j+1,\alpha} - \frac12 \right)\right].
\label{eqXXZlead}
\end{align}   
By subtracting $1/2$ from the occupancy operator $\hat n_{j,\alpha}$ the Hamiltonian is particle-hole
symmetric.   
Employing a Jordan-Wigner transformation (see  e.g. Ref.~\onlinecite{Giamarchi}), it can alternatively 
be written in terms of spin degrees of freedom (XXZ Heisenberg model). 

Two different realizations of the contact between the reservoirs are considered. 
The first one of these is 
referred to as  \emph{point contact} and described by
\begin{align}
\label{eqPointContact}
H_{\rm{coup}}(t)&=\Theta(t)J^\prime \left[\frac{1}{2}c_{1,{\rm L}}^\dagger c_{1,{\rm R}}+\frac{1}{2}c_{1,{\rm R}}^\dagger c_{1,{\rm L}}  \right. \notag \\ &+ \left. \Delta^\prime \left(\hat n_{1,{\rm L}}- \frac12 \right) \left(\hat n_{1,{\rm R}} - \frac{1}{2} \right)\right] .
\end{align}
In this a hopping $J^\prime$ as well as a nearest-neighbor interaction $J^\prime \Delta^\prime$ across 
the link connecting the left and right chains is turned on. 
It is schematically shown in Fig. \ref{fig:setup} (c). 
The second one is called \emph{dot contact} and the corresponding part of the Hamiltonian reads
\begin{align}
\label{eqDotConctact}
H_{\rm{coup}}(t)&= \epsilon \hat n_0 + \Theta(t)\sum\limits_{\alpha={\rm L,R}} J^\prime_\alpha 
\left[\frac{1}{2}c_{1,\alpha}^\dagger c_{0}+\frac{1}{2}c_{0}^\dagger c_{1,\alpha}
  \right. \notag \\ &+ \left. \Delta^\prime_\alpha \left(\hat n_{1,\alpha}- \frac12 \right) \left(\hat n_{0} - \frac{1}{2} \right)\right] .
\end{align}
In this contact, the two reservoirs are tunnel coupled by a single dot site of energy $\epsilon$ 
located at $j=0$; see Fig. \ref{fig:setup} (a). We note that these point and dot contact 
impurity problems are described by lattice systems with total number of sites $2L$ and $2L+1$, 
respectively. 

Additionally, we study the case of a single interacting reservoir 
[$H_{{\rm chain,L}}$ of Eq.~\eqref{eqXXZlead}] tunnel coupled
by a term of the form of Eq.~\eqref{eqDotConctact}, but only considering the $\alpha={\rm L}$ part, 
to a single level quantum dot at $t=0$; see Fig.~\ref{fig:setup} (b). This system is described 
by a lattice of in total $L+1$ sites and referred to as the \emph{single-lead case} in the 
following.  

\emph{Initial preparation} --- The initial density matrix is prepared in each case as a product  
\begin{equation}
\rho=\left| \Psi_{0} \right \rangle \left\langle \Psi_{0 } \right|=\!
\begin{cases}
\rho_{\rm L} \otimes \rho_{\rm R} & {\text{point contact}}\\ 
\rho_{\rm L} \otimes \rho_{\rm{dot}} \otimes \rho_{\rm R}  & {\text{IRLM or dot contact}}\\
\rho_{\rm L} \otimes \rho_{\rm{dot}}   & {\text{single-lead case}}
\end{cases}
\end{equation}
with $\rho_{\alpha}=\left| \Psi_{0,\alpha} \right \rangle \left\langle \Psi_{0,\alpha } \right|$ given 
by the ground states $\left| \Psi_{0,\alpha} \right \rangle$ of the 
reservoirs at half filling. In the examples involving a quantum dot,  $\rho_{\rm{dot}}$ is given 
by the vacuum  (a spin-down in the spin representation).

\subsection{Renormalization group arguments and scaling}

It is well established that the \emph{equilibrium} low-energy physics of both 
our non-interacting  as well as our interacting 1d reservoirs can be described by a 
gapless continuum field theory --- the Tomonaga-Luttinger model --- as long as 
$- J_\alpha < \Delta_\alpha < J_\alpha$ (at half-filling).~\cite{Giamarchi} For $\Delta_\alpha \neq 0$  
our spinless fermion model falls into the Luttinger liquid universality class. 
All terms not captured by the field theory are RG irrelevant and 
flow to zero. However, the bare amplitude of the leading irrelevant bulk couplings 
(e.g.~the umklapp scattering for $\Delta_\alpha >0 $) grows if one approaches the transitions to 
the gapped phases 
$|\Delta_\alpha | \to J_\alpha$. This leads to a decreasing energy scale below which Luttinger Liquid physics can be observed. This energy scale vanishes at the phase transition.~\cite{Karrasch12} 
Thus, if 
the range of accessible energies is bounded from below, e.g. by finite size effects, it might be 
impossible to observe the physics of the field theory.~\cite{Karrasch12,Kennes14a} 
We here avoid this problem by restricting the interaction strength 
to $|\Delta_\alpha | \lesssim 0.7$. 

It was shown that at low energies the equilibrium or steady-state impurity 
physics established when the above local couplings between the reservoirs 
are active for all times is captured by field theory.~\cite{IRLM2,Karrasch10,Enss05}  
We here focus on quench setups where the coupling between the reservoirs, 
be it via a structureless point contact or via a quantum dot, is a relevant perturbation 
in the RG sense leading to a flow to strong coupling. This means that the 
{\em effective} coupling between the reservoirs should be thought of as scale 
dependent; it is growing as the energy scale is lowered across the typical energy  $T_K$. 
The dependence of $T_K$ on the parameters of the Hamiltonian depends on the model 
considered and is given below. This flow can even be found if the bare couplings 
are infinitesimally small. The RG flow from weak to strong coupling is sometimes 
called `healing flow' for obvious reasons. In the presence of a single additional 
energy scale besides $T_K$, e.g.~the temperature $T$, it implies single parameter scaling 
of observables with variable $T/T_K $ and the physics can be regarded as 
{\em universal.} 

This equilibrium physics becomes particularly transparent if we further 
restrict the parameters of our models. To avoid any complications due to the 
interplay of several energy scales in our dot models we assume $\epsilon=0$; 
the dot fulfills the resonance condition. For the IRLM  it was shown that for 
left-right asymmetric setups the relation between the flowing level-reservoir 
hoppings and observables becomes involved.~\cite{Karrasch10a} As a consequence the 
latter no longer shows the simple scaling behavior found 
for restored left-right symmetry. In Luttinger liquids the sharp crossover between two 
half chains with different bulk interactions effectively induces single-particle scattering 
at the contact and destroys the resonance condition.~\cite{Enss05} Although such asymmetry 
effects are of genuine interest and of relevance in most experimental realizations, in 
our first attempt to understand the non-equilibrium dynamics we 
suppress them by considering left-right symmetric setups with 
$U_\alpha= U$ and $\gamma_\alpha^{(\prime)}=\gamma^{(\prime)}$ for the IRLM as 
well as  $J^{(\prime)}_\alpha=J^{(\prime)}$ and $\Delta^{(\prime)}_\alpha=\Delta^{(\prime)}$ 
for the models with Luttinger liquid leads. We furthermore fix our unit of energy by setting 
$\gamma=1=J$.  

\begin{figure}[tb]
 \centering
 \includegraphics[clip,width=0.45\textwidth]{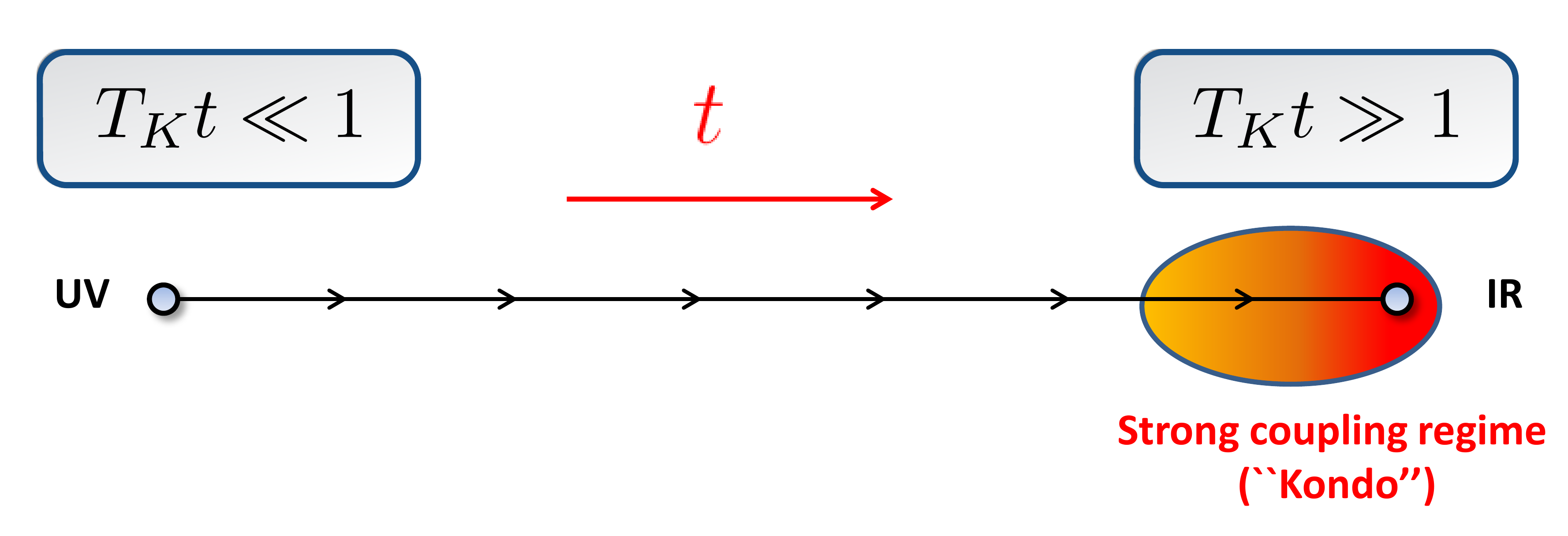}
\caption{(Color online)  Dynamical healing RG flow studied in this work. The dynamics only depends on $T_K t$, such that the time evolution follows the RG flow from weak to strong coupling. In particular, the large time limit 
$ T_K t \gg 1$ corresponds to the low energy, strong coupling regime. }
  \label{fig:FigRG}
\end{figure}

The question whether the same type of universality is realized in the 
non-equilibrium post-quench dynamics is largely open; it lies at the heart of our 
present work. More specifically we investigate numerically whether the entanglement 
entropy and the Loschmidt echo introduced below are scaling functions 
with $T_K t$ being the relevant variable indicating a `dynamical healing'. We thus 
ask  whether `time follows the RG flow', with the large time 
limit $T_K t \gg 1$ corresponding to the low energy fixed point (see Fig.~\ref{fig:FigRG}
for a sketch). This is what one expects based on a field-theoretical approach to the type of local 
quenches considered here.~\cite{LetterRLM,CardyCalabrese2} In this approach, the 
1d non-equilibrium problem is formally `folded' into a two-dimensional (2d) boundary statistical 
mechanics problem in equilibrium using a Wick rotation.  However, this 
mapping is based on {\em assumptions} on the analytic properties of the appearing partition 
functions. By providing indications of one-parameter scaling with $T_K t$ we thus not only establish 
evidence that the dynamics of the non-equilibrium problems considered here is (1) universal and that (2) 
field theory can be applied but also that (3) the assumptions underlying the field theory approach 
itself are justified. We furthermore compare the long time dynamics of our observables to the 
corresponding field-theoretical predictions. The alleged simple replacement 
of equilibrium energy scales such as temperature by inverse time in standard RG arguments 
is obviously very specific to the considered type of {\em local} quenches. Within field
theory on a technical level it is linked to the possibility of  `folding' and the subsequent 
use of BCFT.    

We already emphasize at this point that universal dynamics can only be expected for times 
$B^{-1} \ll t \ll L/v_F$ with $B$ being the reservoirs band width and $v_F$ their Fermi
velocity. This holds in close analogy to the equilibrium situation in which scaling is cut off 
at large energies by non-universal band effects and at low ones by finite size effects.

\subsection{Observables}

There are many different observables that could be considered to monitor the dynamics 
of the system after the quantum quench. A very instructive quantity is the entanglement of the 
sub-systems as a function of time that we measure by the entanglement entropy $S(t)$. 
It characterizes the exchange of information between the two reservoirs.
More precisely, the entanglement entropy is calculated between the left reservoir (${\rm L}$) 
and the rest of the system (${\rm \bar L}$ which in our dot setups contains the dot site) 
and is defined as
\begin{equation}
S(t)=-\rm{Tr}\left\{\rho_{\rm L}(t){\rm{ln}}\left[\rho_{\rm L}(t)\right]\right\},
\end{equation}
with the reduced density matrix of the left subsystem $\rho_{\rm L}(t)={\rm{Tr}}_{\rm \bar L}
\left[\rho(t)\right]$ with respect to the full system $\rho(t)=\left| \Psi(t) \right \rangle\left\langle \Psi(t)
 \right|$ and $\left| \Psi(t) \right \rangle=e^{-iHt}\left| \Psi_0 \right \rangle$.
Initially, we have $S(0)=0$ as the left reservoir is decoupled from the rest of 
the system. The evolution of the entanglement entropy across a marginal defect in non-interacting 
fermionic systems was computed in Refs.~\onlinecite{EE1,EE2,EE3}. The generalization to interacting problems 
in the case of a weakly modified link was carried out recently in 
Ref.~\onlinecite{CalabreseWeakLink} (see Ref.~\onlinecite{EE4} for the corresponding analysis in equilibrium). 
In the following, we investigate whether the results of Ref.~\onlinecite{CalabreseWeakLink} hold 
for a microscopic model. In addition, we study $S(t)$ for dot setups. 

Another quantity that we study is the so-called Loschmidt echo
\begin{equation}
G(t)=\left\langle \Psi_0 \right|e^{iH_0 t} e^{-iH t} \left| \Psi_0 \right \rangle,
\end{equation}
which is the Fourier transform of the work distribution~\cite{WorkSilva1} 
$
P(W)=\sum_n\delta\left(W-\left[E^{(n)}-E_0 \right]\right)\left|\big\langle \Psi^{(n)}
\big| \Psi_0\big\rangle\right|^2,
$
with the eigenvalues $E^{(n)}$ and eigenstates $\big| \Psi^{(n)}\big\rangle$ of the Hamiltonian 
after the quench, and $\big| \Psi_0\big\rangle$ the ground state  
of the Hamiltonian before the quench (decoupled reservoirs) with 
corresponding ground-state energy $E_0$.
The square modulus of the Loschmidt echo is the probability to find 
the system at time $t$ in its initial state. It can be thought of as a quantum return probability
which, if decaying, characterizes the irreversibility of the time evolution. The Loschmidt echo has 
attracted a lot of attention recently, in the context of `dynamical' Anderson orthogonality 
catastrophes, see e.g.~Refs.~\onlinecite{KondoQuench1,KondoQuench2,EchoCFT2, AOCrec1,AOCrec2,AOCrec3}, but 
also in studies of `dynamical phase transitions'.~\cite{DynamicalPhasetranstion,DynamicalPhasetranstion2,
DynamicalPhasetranstion3,DynamicalPhasetranstion4,DynamicalPhasetranstion5,DynamicalPhasetranstion6} 
We note that the work distribution can be argued to be directly proportional to the absorption 
spectrum in quantum dot setups where the absorption of a photon effectively triggers a quantum 
quench.~\cite{KondoQuench1,KondoQuench2,HKXray} A general field theoretical formalism to compute 
$G(t)$ for (`integrable') impurity problems after a local quench was developed in Ref.~\onlinecite{LetterRLM} but only 
applied to the non-interacting resonant level model (RLM).

\subsection{Methods}

\emph{Boundary Conformal Field Theory} ---
In general, it is difficult to provide closed analytical expressions for the time evolution of observables 
after a local quench. Within field theory this can be linked to the emergence of an 
energy scale $T_K$ that breaks conformal 
invariance, and to the presence of interactions. However, in the large time limit $ T_K t \gg 1$, one can  
obtain asymptotic results using the above described mapping 
to a 2d inhomogeneous statistical mechanics problem as well as BCFT. 
For the entanglement entropy one finds~\cite{CardyCalabrese2,CardyCalabrese1} 
\begin{equation}
S(t) \underset{T_K t \gg 1}{\longrightarrow} S_0+\frac{1}{3} \log (T_K t), 
\label{eqCFTEntropy}
\end{equation}
where we have inserted the central charge $c=1$ of the Luttinger liquid (or Dirac fermions) 
reservoirs (see e.g. Ref.~\onlinecite{Giamarchi}). 
Here $S_0$ is a non-universal constant. 
This relation should hold in all setups with two reservoirs --- in the single-lead case, the 
entanglement entropy is bounded from above by $\ln 2$ (see below).

Computing the large time asymptotics of the Loschmidt echo is slightly more involved. For a {\em point contact 
junction} between two Luttinger liquids, we can use the CFT results of Refs.~\onlinecite{EchoCFT1,EchoCFT2} and obtain
\begin{equation}
|G (t)|^2 \underset{T_K t \gg 1 }{\sim}  \left(T_K t \right)^{-1/4},
\label{eqLoschmidtpointcontact} 
\end{equation}
independent of $\Delta$ as long as $-1< \Delta < 0$ so that the local tunnel coupling 
is relevant. Note 
that we have once again inserted the central charge $c=1$ of the gapless reservoirs. 

The {\em IRLM} and {\em dot setups} are 
more subtle because of the dynamical nature of the impurity. In these cases, it is convenient to use the general 
framework introduced in Ref.~\onlinecite{LetterRLM} and consider the imaginary time Loschmidt echo as a correlation 
function (or modified partition function) in a semi-infinite critical 2d statistical mechanics 
problem, with the impurity corresponding to a boundary condition; see Ref.~\onlinecite{AffleckLudwigFermiEdge} for 
a similar calculation in the context of the Fermi edge singularity. In that language, a quantum quench at 
$t=0$ can be considered as a sudden change of boundary condition at imaginary time $\tau=0$, effectively 
creating infinitely many massless excitations in the bulk. Following Ref.~\onlinecite{LetterRLM}, one can then 
interpret the Loschmidt echo $G(t=-i \tau)$ for $T_K \tau \gg 1$ as the two-point function of a boundary condition changing 
operator,~\cite{Cardy} whose scaling dimension is fixed by conformal invariance. Assuming analyticity to rotate 
back to real time, we expect~\cite{LetterRLM}
\begin{equation}
|G(t)|^2 \underset{T_K t \gg 1 }{\sim}  \left(T_K t\right)^{- 4 \Delta_{\rm BCC}}, 
\label{eqGeneralFormulaLoschmit}
\end{equation}
with the scaling dimension $\Delta_{\rm BCC}$ which is  model dependent and given below. We note that 
the vanishing of the Loschmidt echo at large time is the real-time analog of the well-known Anderson 
orthogonality catastrophe.~\cite{Anderson} As a consequence the work distribution is characterized by an edge 
singularity at low energy~\cite{WorkSilva1, HKXray,WorkSilva2} which can be observed in optical absorption 
experiments (see e.g.~Refs.~\onlinecite{KondoQuench1,KondoQuench2} in the context of the Kondo effect).

\emph{Density Matrix Renormalization Group} --- 
DMRG has proven to be an invaluable tool to numerically study the equilibrium and non-equilibrium many-body physics 
of  interacting one-dimensional systems.~\cite{Schollwoeck11} In DMRG the numerical cost depends in an 
exponential fashion on the entanglement in the system. For typical real-time evolutions, 
entanglement grows linearly with time and thus the numerical resources available are exhausted 
exponentially fast until no further progress in time can be made. The time scales reachable within DMRG 
thus critically depend on the entanglement of the system under scrutiny.  In this work, we apply DMRG in a 
very natural representation via matrix product states.~\cite{Schollwoeck11,Vidal04,White04} We use the 
implementation outlined in Ref.~\onlinecite{Schollwoeck11} to tackle the problems introduced in the previous 
section. We checked that preparing the ground states in the decoupled reservoirs either via a simple 
imaginary time evolution starting from an initial random product state (see Sect.~7 of Ref.~\onlinecite{Schollwoeck11}) 
or a more sophisticated iterative procedure (see Sect.~6 of Ref.~\onlinecite{Schollwoeck11}) yield coinciding 
results. The iterative ground state search is performed with a single-site algorithm. To protect this 
algorithm from getting stuck in non-global minima when optimizing the energy we use the ideas 
introduced in Ref.~\onlinecite{White05}. For the imaginary time evolution, we apply a second-order Trotter 
decomposition with $\Delta \tau = 0.0025$, and we gradually increase the bond dimension $\chi$ during 
the convergence process. Once the ground states have been prepared, we employ a real time evolution 
algorithm (see again Sect.~7 of Ref.~\onlinecite{Schollwoeck11}) with the full Hamiltonian including the 
tunneling terms. We use a fourth-order Suzuki-Trotter decomposition
with $\Delta t =0.2 $ chosen small enough to give converged results on the scale of every plot presented 
in the following. Additionally, we can exploit the trivial rewriting of the Loschmidt 
echo \cite{Barthel13,Kennes14}
\begin{align}
G(t)&=\left\langle \Psi_0 \right|e^{iH_0 t} e^{-iH t} \left| \Psi_0 \right \rangle \notag \\ &=e^{iE_0 t}\left\langle \Psi_0 \right|
e^{-iH t/2} e^{-iH t/2} \left| \Psi_0 \right \rangle  \notag \\ &=e^{iE_0 t}\left\langle \Psi_0(-t/2) | \Psi_0(t/2) \right \rangle .
\end{align}
This allows us to reach times twice as large as in the original form exploiting the same numerical 
resources. The bond dimension is dynamically increased during the real time evolution so that the discarded 
weight $\epsilon$ always remains below $10^{-7} - 10^{-8}$.
 
\emph{Functional Renormalization Group} --- For the IRLM we also analyze the quench dynamics 
using the FRG. FRG is a versatile method to tackle 
quantum many-body problems.~\cite{Metzner12} Recently, it was extended to time evolution in
non-equilibrium \cite{Kennes12} including quench dynamics.~\cite{Kennes12a,Kennes13}  
In the present case it allows to study our lattice realization of the IRLM at larger system sizes as well as to 
reach larger times compared to DMRG.  

In FRG one sets up an infinite hierarchy of flow equations for the many-particle vertex functions.  
We here employ the lowest order truncation scheme for this hierarchy. This means that the two-particle 
vertex is set constant (remaining the bare two-particle interaction throughout the entire flow), while in 
contrast the self-energy acquires a RG flow. This truncation poses the only relevant approximation to the 
many-body problem at hand. It is known that this procedure is well suited to describe the impurity physics of 
the IRLM. It leads to the correct resumation of logarithmic terms in form of power laws,  with 
exponents agreeing to the exact ones to leading order in $U$.~\cite{Karrasch10,Kennes12}

We supplement the scheme of Ref.~\onlinecite{Kennes13} by a Suzuki-Trotter decomposition as used in DMRG to describe 
the propagation in time. This significantly boosts the performance. We always choose a symmetric fourth order 
decomposition, with $\Delta t=0.1$ taken small enough such that it corresponds to a negligible approximation. 

Since the FRG scheme outlined in Ref.~\onlinecite{Kennes13} aims at the single particle Green functions we need to 
explain how the Loschmidt echo can be deduced. In our truncation order interactions are incorporated 
effectively in a non-interacting, but time-dependent, renormalized Hamiltonian (in form of the self-energy). 
Since the time steps $\Delta t=0.1$ used are small enough, such that the Hamiltonian can be approximated 
as a constant $H_t$ during each of such time steps we can straightforwardly perform the time evolution of 
the initial wave function $\left|\Psi(0)\right\rangle$ (ground state of the decoupled reservoirs) with 
the renormalized Hamiltonian. More explicitly, we can write the ground state of the non-interacting 
reservoirs as a product state \cite{Rigol05}
\begin{equation}
\left|\Psi(0)\right\rangle=\prod\limits_{m=1}^{N_f}\left(
\sum\limits_{j=1}^LP_{jm}(0)c_j^\dagger\right)\left|0\right\rangle,
\end{equation}
where $N_f$ denotes the total number of fermions and $\left|0\right\rangle$ is the 
vacuum. The wave function can be propagated in time with a stepwise (in time) constant 
Hamiltonian $H_t=\sum\limits_{i,j}\epsilon_{t,ij}c_i^\dagger c_j$ by
\begin{equation}
\begin{split}
\left|\Psi(t+\Delta t)\right\rangle&=e^{-i H_t\Delta t}\left|\Psi(t)\right\rangle
 \notag \\ &
 = \prod\limits_{m=1}^{N_f}\left(\sum\limits_{j=1}^LP_{jm}(t+\Delta t)c_j^\dagger\right)\left|0\right\rangle , \\
\underline P(t+\Delta t)&=e^{-i\underline H_t\Delta t} \underline P(t),
\end{split}
\end{equation}
where $\underline P$ and $\underline H_t$ are the matrices with entries $P_{jm}$ and $\epsilon_{t,jm}$, respectively.
The Loschmidt echo in turn can then be calculated as
\begin{equation}
\begin{split}
&G(t)=\left\langle\Psi(0)\right|e^{-iH_0t}e^{-iHt}\left|\Psi(0)\right\rangle 
\\&=e^{-iE_0t}\left\langle\Psi(-t/2)|\Psi(t/2)\right\rangle\\
&=e^{-iE_0t}\left\langle 0 \right|\prod\limits_{m=1}^{N_f}
\sum\limits_{j=1}^LP_{jm}(-t/2)^*c_j\prod\limits_{n=1}^{N_f}\sum\limits_{i=1}^LP_{in}(t/2) c_i^\dagger\left|0\right\rangle\\
&=e^{-iE_0t} {\rm{det}}\left[\underline{P}^\dagger(-t/2) \underline{P}(t/2)\right].
\end{split}
\end{equation}
While in the outlined truncated FRG treatment single-particle Green functions are approximated correctly (at least) to 
leading order in $U$ this is less clear for a quantity like the Loschmidt echo. We have therefore carefully 
checked our FRG results against numerically exact DMRG data for shorter times and smaller systems 
(see Fig.~\ref{fig:IRLM_com_FRG_DMRG}). The agreement at small $U$ is very convincing such that we can trust 
the FRG results for $G(t)$ at larger $L$ and $t$ as well. In a similar, but more complicated fashion one could 
derive an  expression for $S(t)$ accessible to FRG; we do not pursue this here for the sake of brevity.
   
\section{The interacting resonant level model}

\subsection{Field theory limit}
\label{subsecFTIRLM}

In the field theory description of the semi-infinite chain-like reservoirs, we first linearize the single-particle dispersion 
near the Fermi energy and then conveniently ``unfold'' these half-infinite wires  to obtain infinite ones with only right moving fermions. This leads to  
$H_{\rm{res},\alpha}=- i v_F \int_{-\infty}^\infty d x \psi_\alpha^\dagger(x) \partial_x \psi_\alpha(x)$, 
with the Fermi velocity $v_F=2$ and the field operators $ \psi_\alpha^{(\dagger)}(x)$. 
The coupling between the dot and the reservoirs can then be expressed 
as $H_{\rm{coup},\alpha}(t) = \tilde{\gamma}^\prime \Theta(t) \left( c^\dagger_0 \psi_\alpha(0)
+\psi^\dagger_\alpha(0) c_0 \right)+\tilde{U} \left(\hat n_0- \frac12 \right) :\psi^\dagger_\alpha(0) \psi_\alpha(0): $, 
where $\tilde{\gamma}^\prime$ and $\tilde{U}$ depend on the microscopic parameters $\gamma^\prime$ and $U$ 
in a non-universal (and in general unknown) way. Here $: \ldots :$ denotes normal ordering. We then 
bosonize the fermionic fields $\psi_\alpha \sim \mathrm{e}^{i \sqrt{4 \pi} \phi_\alpha}$,~\cite{Giamarchi}
with bosonic fields $\phi_\alpha(x)$, and introduce an effective spin operator through $c^\dagger_0 = \eta S^+$ 
and $\hat n_0 = S^z+\frac{1}{2}$ with $\eta$ a Majorana fermion. After several canonical transformations, 
the IRLM Hamiltonian can be mapped onto an effective anisotropic Kondo problem,~\cite{IRLM1} described by a single 
chiral boson $\phi(x)$ that depends on $\phi_{\alpha={\rm R}/{\rm L}}$ in a complicated non-linear fashion (see e.g. Ref.~\onlinecite{IRLM1} for details of the mapping). 
The resulting Hamiltonian reads
\begin{equation}
H = \int_{-\infty}^\infty dx 
[\partial_x \phi(x)]^2 +\Theta(t)  \frac{\tilde{\gamma}^\prime}{\sqrt{\pi}} 
\left[\mathrm{e}^{i \beta \phi(0)} S^+ + {\rm H.c.} \right],
\label{eqIRLMFT}
\end{equation}
where the boundary perturbation (second term) has dimension $h = \frac{\beta^2}{8 \pi} = \frac{1}{4 \pi^2} 
(\tilde{U}- \pi )^2 + \frac{1}{4} $ (assumed to be smaller than one). This means that the boundary perturbation 
flows under a RG procedure as
\begin{equation}
\frac{{\rm d} \tilde{\gamma}'}{{\rm d} \ell} = (1-h)  \tilde{\gamma}' + \dots 
\end{equation}
with $\ell = \ln \Lambda$, and $\Lambda$ is the infrared cutoff. Solving this equation leads to the length 
dependent coupling $ \tilde{\gamma}'(L)  \simeq   \tilde{\gamma}' L^{1-h} $, where  $\tilde{\gamma}'$ is 
the bare coupling, and $L$ a typical length scale. The Kondo temperature $T_K \sim \frac{1}{L_K}$ is defined as the scale at which  
$\tilde{\gamma}'(L_K) \simeq 1$. It thus scales as
\begin{equation}
T_K  \propto (\tilde{\gamma}')^{1/(1-h)} \propto (\gamma')^{1/(1-h)}.
\end{equation}
The RLM has $U=\tilde{U}=0$ such that $h=\frac{1}{2}$ and $T_K \propto (\gamma')^2$. For 
small $U$, one finds $h = \frac{1}{2}-\frac{U}{2\pi} +  {\mathcal O}(U^2)$ (see e.g.~Ref.~\onlinecite{Doyon}). 
A particularly interesting value 
of the Coulomb interaction is given by the so-called self-dual point for which $\tilde{U}=\pi$ in our 
regularization scheme, such that $h=\frac{1}{4}$ and $T_K \propto (\gamma')^{4/3}$. This self-dual point 
corresponds to $U \approx 2$ on the lattice.~\cite{IRLM2} Without loss of generality we fix the prefactor 
of $T_K$ to be 4. Why this choice is useful will become clear in the next section.

At low energy, the boundary interaction flows to the conformally invariant boundary condition 
$\phi(0^+)=\phi(0^-)+\delta/\sqrt{\pi}$, where $\delta^2 =\pi^2 h/ 2$  corresponds to the phase shift 
felt by the effective fermion $\psi \sim \mathrm{e}^{i \sqrt{4 \pi} \phi}$. 
We stress that this fermion $\psi$ has a very complicated expression in terms of the original fermionic 
fields $\psi_{\alpha={\rm L,R}}$. As discussed in Ref.~\onlinecite{LetterRLM} the large time behavior of the entropy is 
given by Eq.~\eqref{eqCFTEntropy}, whereas the Loschmidt echo behaves as 
Eq.~\eqref{eqGeneralFormulaLoschmit} with 
\begin{equation}
\Delta_{\rm BCC}=\frac{1}{2} \left(\frac{\delta}{\pi}\right)^2=\frac{h}{4} .
\end{equation}
Using Eq.~\eqref{eqGeneralFormulaLoschmit} we therefore expect $|G(t)|^2 \sim (T_Kt)^{-1/2}$ for the RLM and 
$|G(t)|^2 \sim (T_Kt)^{-1/4}$ for the IRLM at the self dual point. This provides an interesting example where 
the interactions strongly renormalizes the large time exponent of the Loschmidt echo.

\subsection{Microscopic model and scaling limit}
\label{subsecScalinglimit}

\begin{figure}[tb]
 \centering
 \includegraphics[width=0.44\textwidth]{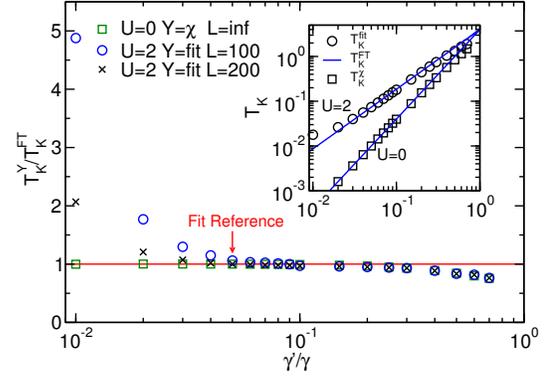} 
 \caption{(Color online) Dependence of the Kondo temperature on $\gamma'$
for the IRLM (including the RLM with $U=0$). We show the ratio of 
 the Kondo temperature $T^{\rm fit}_K$ obtained by collapsing 
 our numerical DMRG data by hand, the $T^{\chi}_K$ from Eq.~\eqref{TKchi}, 
 and the field theory prediction $T_K^{\rm FT} = 4 (\gamma')^{\frac{1}{1-h}}$.}
  \label{fig:Tkmicro}
\end{figure}
\begin{figure*}[tb] 
 \centering
 \includegraphics[width=0.48\textwidth]{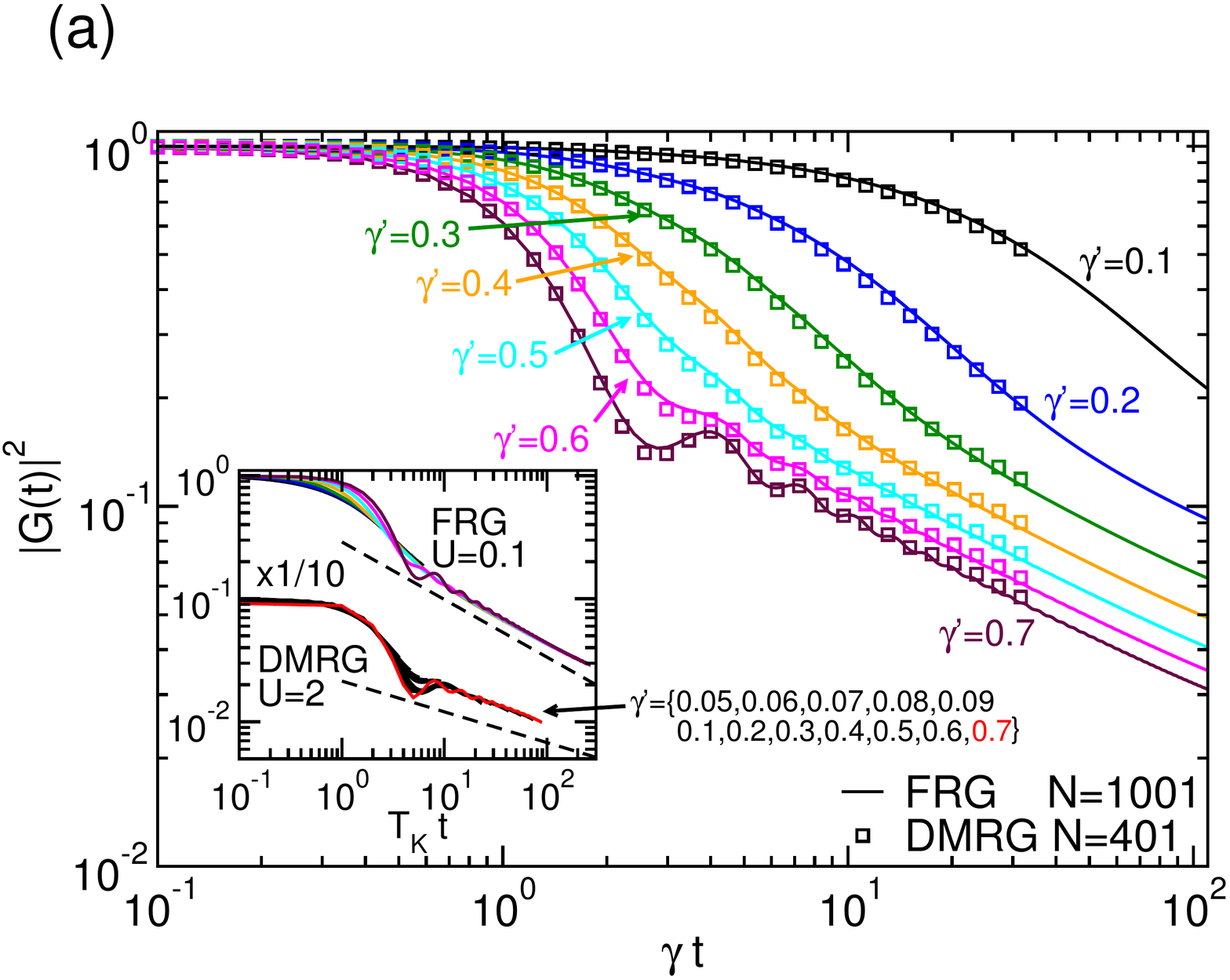}
\hspace{0.1cm}
 \includegraphics[width=0.48\textwidth]{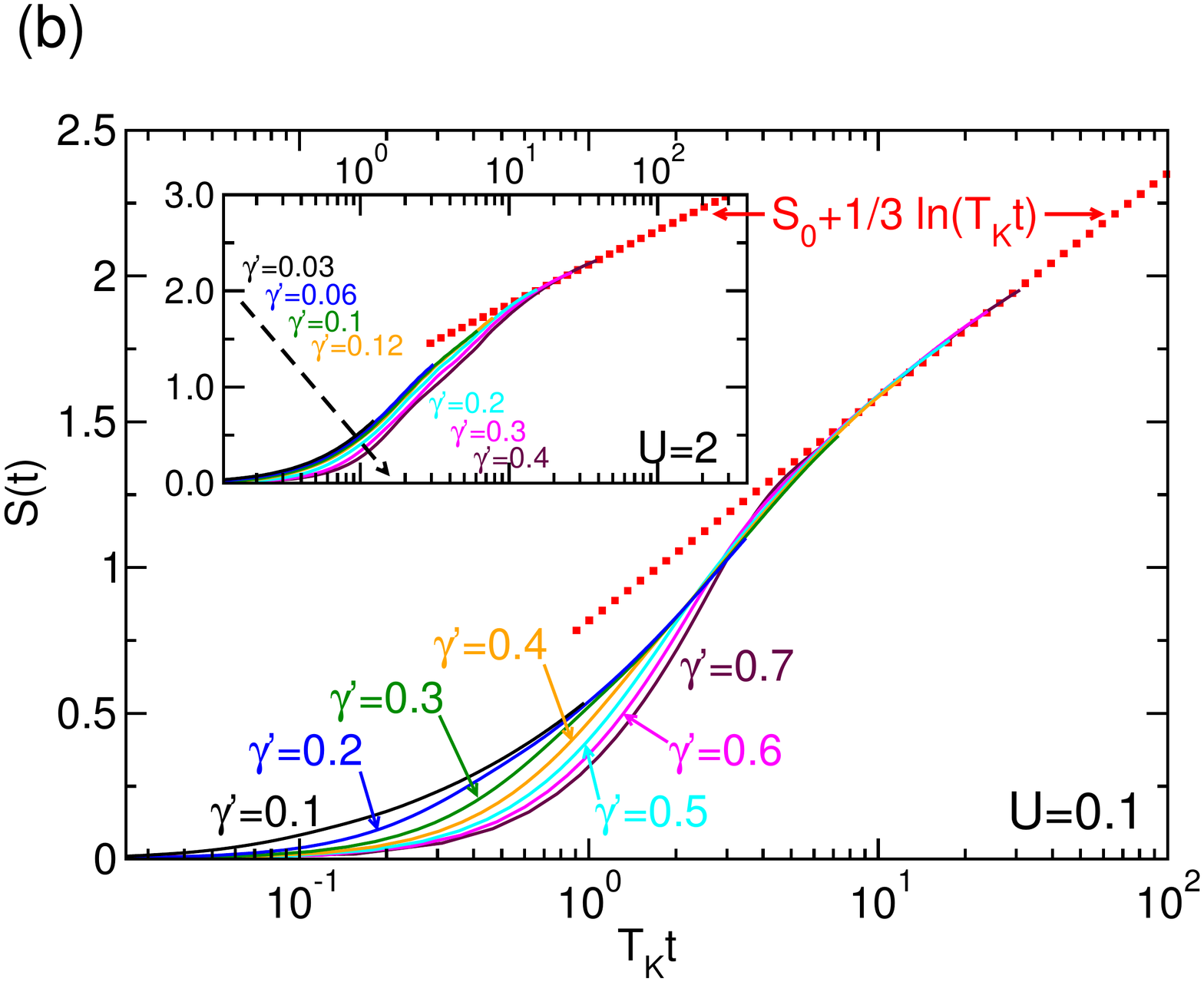}
 \caption{(Color online) Time evolution of the Loschmidt echo and the entanglement entropy after a quantum 
 quench in the IRLM. (a) Comparison between our FRG and DMRG data for the Loschmidt echo at $U=0.1$. 
Inset: universal collapse of the Loschmidt echo curves when rescaled 
 by the Kondo temperature $T_K$ (for $U=0.1$ and $U=2$). The dotted lines correspond to 
the large time BCFT predictions. 
 (b) Universal scaling of the entanglement entropy for $U=0.1$ and $U=2.0$ (inset). At large 
 time, one recovers the expected CFT behavior (red dotted line).}
  \label{fig:IRLM_com_FRG_DMRG}
\end{figure*}
When comparing predictions from field theory to results obtained numerically for the microscopic models 
obtained by DMRG and FRG, one needs to be careful about the range of validity of the 
field theory description. Two conditions have to be fulfilled simultaneously: (a) 
the large $L$ limit $T_K L/v_F \gg 1$ and (b) the scaling limit $T_K / B \ll 1$. At a given large $L$ 
they impose an upper as well as lower bound on $\gamma'$. As a first step let 
us assume correspondence between the microscopic model and the field theory in the proper limit and let 
us choose $\gamma'=0.05$ with $L=200$ as a point of reference. 
The corresponding reference curve for the Loschmidt echo $|G(t)|^2$ obtained  for $U=2$ using DMRG is 
scaled with the field theory Kondo temperature $T_K^{\rm{FT}}= 4(\gamma^\prime)^{4/3}$. We then determine the Kondo 
temperature $T_K^{\rm{fit}}$ for the microscopic IRLM by collapsing all curves (with different $\gamma'$) on 
top of this scaled reference curve. The ratio between $T_K^{\rm{fit}}$ extracted by this procedure 
and the field theory prediction $T_K^{\rm{FT}}$ is shown in the main panel of Fig.~\ref{fig:Tkmicro}. For 
$\gamma' \gtrsim 0.3$, we leave the scaling limit regime and deviations between  the field theory 
$T_K^{\rm{FT}}$ and the $T_K^{\rm{fit}}$ of the microscopic model become prominent. For very small 
$\gamma'$, one does no longer fulfill $T_K L/v_F \gg 1$ and finite size effects yield deviations in the 
Kondo temperatures. This is confirmed by the fact that reducing the size of the system 
(from $L=200$ to $L=100$) enhances this effect. 

We remark that for the IRLM it has been shown \cite{Andergassen,Kennes12,Karrasch10,Anders1,Anders2} 
that yet another reasonable definition of $T_K$ at small values of $U/\gamma$ is the charge susceptibility 
\begin{equation}
\label{TKchi}
T_K^{\chi}=-\frac{2}{\pi \chi}\;\;\;\;\;\;\chi= \left.\frac{d\left\langle n\right \rangle}{d\epsilon}
\right|_{\epsilon=0}.
\end{equation}     
This choice of $T_K$ for the microscopic IRLM again results in a very convincing collapse of the curves at 
small $U/\gamma$. With this definition of $T_K$ we can fix the prefactor of the field theory $T_K$ to 4, 
by demanding that the two definitions agree at $U=0$. Since $T_K^{\chi}$ does not rely on calculating the 
Loschmidt echo, we can now for $U=0$ compare $T_K^{\chi}$ with $T_K^{\rm{FT}}$ in the limit $L\to \infty$. 
The condition $T_K L / v_F \gg 1$ is thus guaranteed to hold and the finite size deviations for small $\gamma'$ 
disappear, while effects arising from the second inequality $T_K / B\ll 1$ being no longer fulfilled 
remain roughly the same as for $U=2$ (see Fig.~\ref{fig:Tkmicro}). 

Since overall the Kondo temperatures $T_K$ defined from field theory, or via the fit procedure as well as 
the charge susceptibility for the microscopic model agree reasonably well, in the following we will only 
use the field theory definition $T_K=T_K^{\rm{FT}}$ to collapse the curves obtained numerically for the 
microscopic model. Nevertheless, it is important to keep in mind that simultaneously fulfilling both inequalities 
in numerical computations for finite chains infers restrictions on $\gamma'$.

\begin{figure}[bt]
 \centering
 \includegraphics[width=0.44\textwidth,clip]{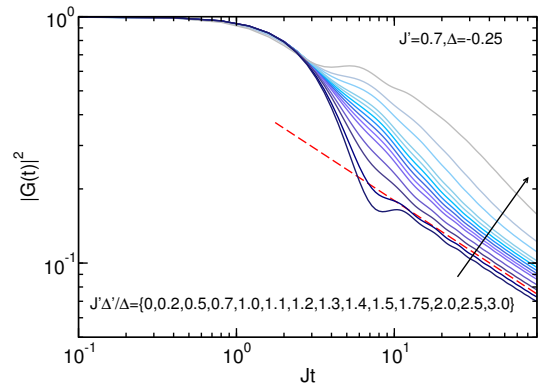}
 \caption{(Color online)  DMRG data for $G(t)$ in the single lead case at fixed $J'$.  
Gradually increasing the interaction $J'\Delta'$ between the dot and the single lead ($L=200$) 
reveals that the marginal term dropped in our field theory analysis does not alter the 
large time exponent. The time scales for which asymptotic behavior can be observed, however, 
increases with increasing $J'\Delta'$. The field theory prediction $|G(t)|^2\sim (T_K t)^{-\frac{1}{2g}}$ is 
included as a dashed red line.}
  \label{fig:Singlesite_infDL}
\end{figure}
\begin{figure*}[tb] 

 \centering
 \includegraphics[width=0.49\textwidth]{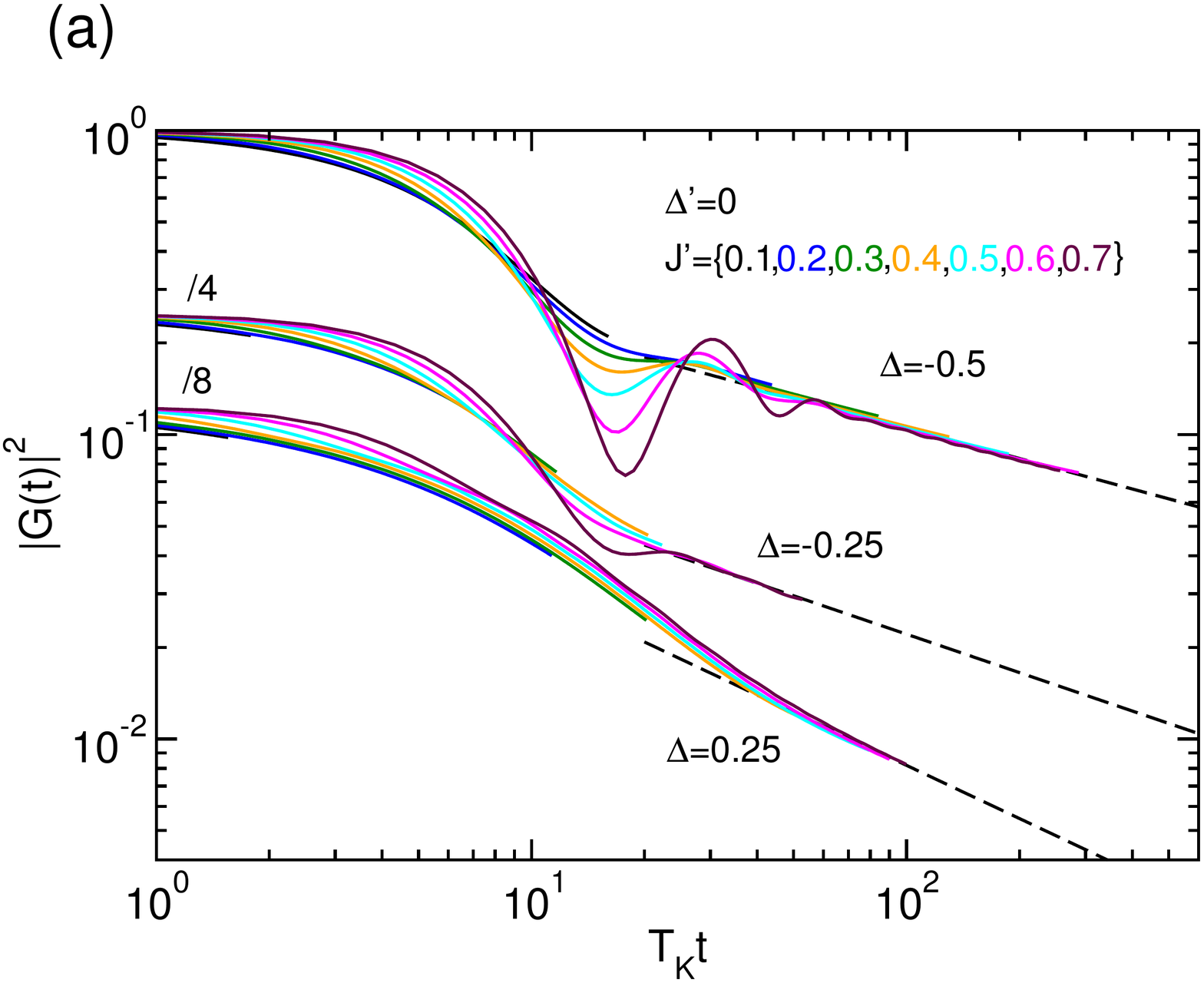}
 \includegraphics[width=0.49\textwidth]{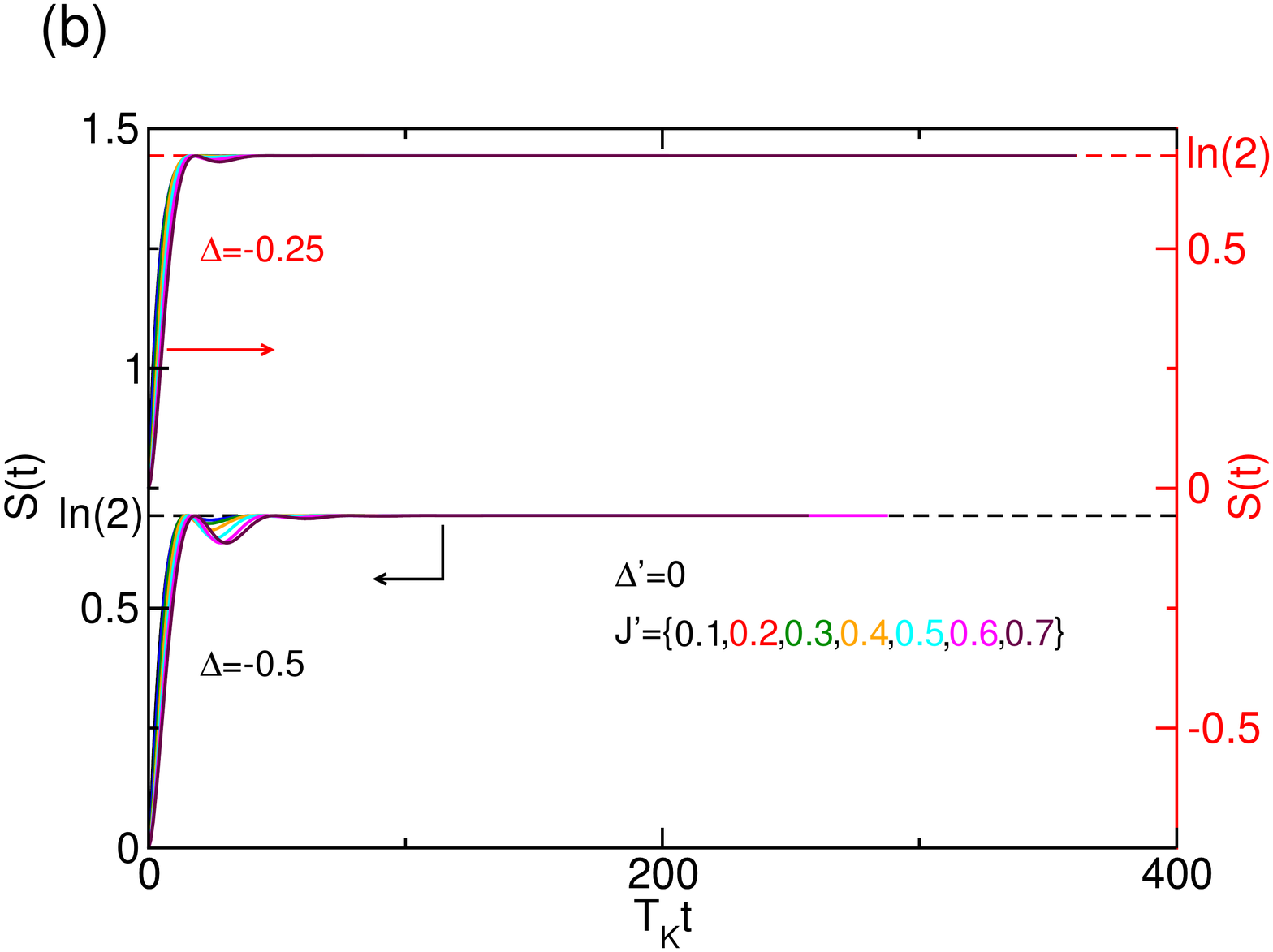}
 \includegraphics[width=0.49\textwidth]{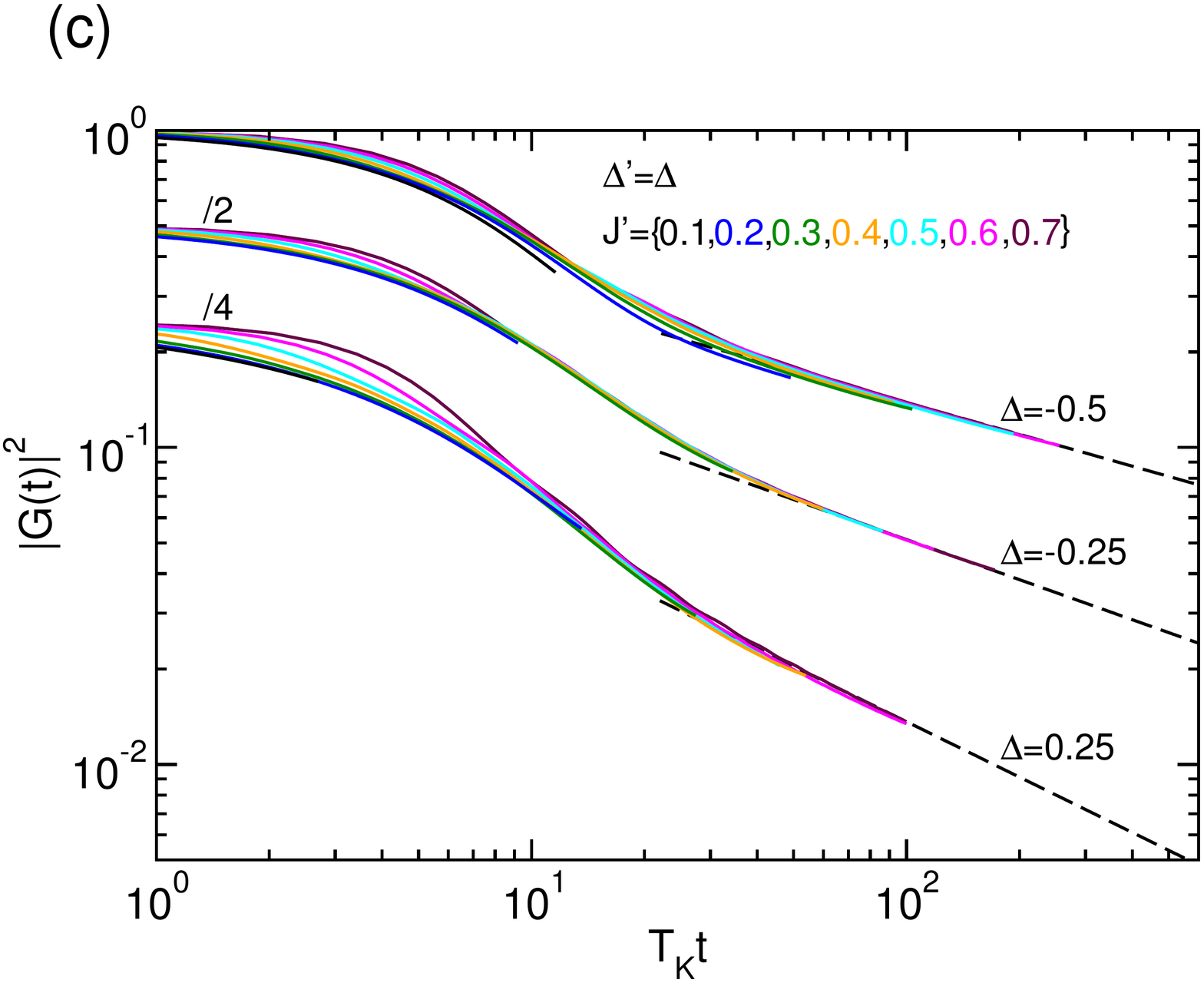}
 \includegraphics[width=0.49\textwidth]{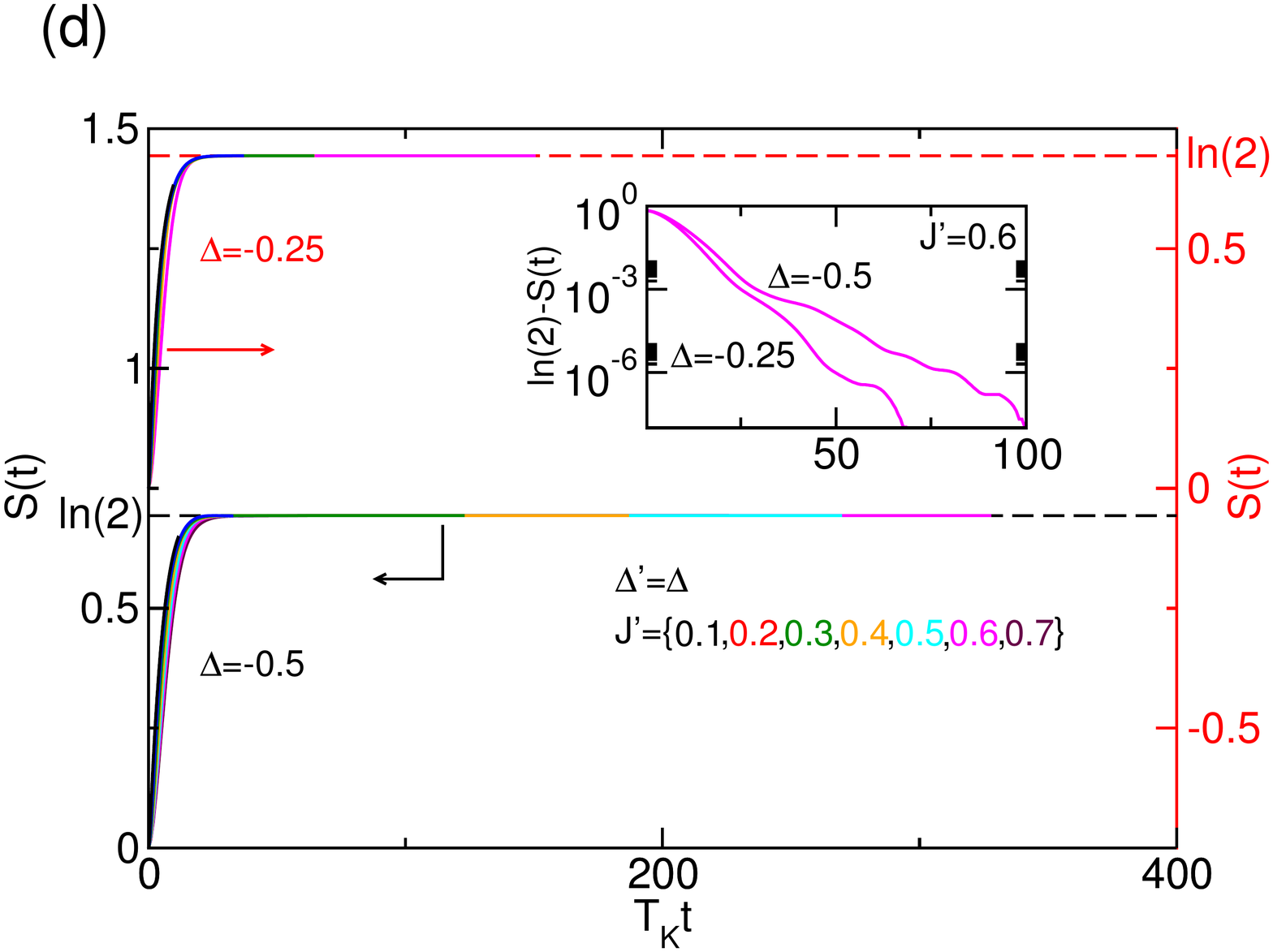}
 \caption{(Color online) Universal scaling of the Loschmidt echo and the entanglement entropy after a quantum quench of the 
tunneling between a Luttinger liquid lead ($L=400$) and a quantum dot. The predicted asymptotic long time 
behavior is indicated as dashed lines. (a), (b) $\Delta'=0$. (c), (d) $\Delta'=\Delta$. (a), (c) 
Loschmidt echo. (b), (d) Entanglement entropy. The inset in (d) shows that the steady-state 
value of $\ln(2)$ is approached exponentially fast, thus indicating that the impurity becomes 
maximally entangled with the lead for large times.}
  \label{fig:Singlesite}
\end{figure*}
\subsection{Quench dynamics of the lattice model}

Let us now discuss the numerical data obtained for the time evolution of the Loschmidt echo as well as the entanglement 
entropy. The main plot of Fig.~\ref{fig:IRLM_com_FRG_DMRG} (a) shows a comparison of the Loschmidt echo obtained 
by DMRG and FRG at $U =0.1$ and for different $\gamma'$. The agreement between the two methods is very convincing for this 
small value of the two-particle interaction. FRG can be used to tackle larger system sizes and times, which is the 
reason why data for $L=500$ is shown compared to $L=200$ for DMRG. The inset of this panel shows the collapse 
of the same FRG curves when the Kondo temperature $T_K$ is used as the proper energy scale. In addition, a similar 
collapse of DMRG data for a larger value of the interaction strength $U=2$ corresponding to the self-dual 
point is presented. The large time BCFT predictions Eq.~\eqref{eqGeneralFormulaLoschmit} of 
Sect.~\ref{subsecFTIRLM} with $\Delta_{\rm BCC}(U=0.1) \approx 0.1210$ and $\Delta_{\rm BCC}(U=2)=1/16 $ 
(shown as dashed lines) are consistent with our numerical data. Similarly, 
Fig.~\ref{fig:IRLM_com_FRG_DMRG} (b) shows the one-parameter scaling of the entanglement entropy. The 
main plot depicts $U=0.1$ while the inset covers $U=2$. The curves collapse well when rescaled 
by $T_K$, and the long time behavior is consistent with the universal CFT prediction for 
the entropy Eq.~\eqref{eqCFTEntropy}.

Considerations about the limits of applicability of field theory scaling in lattice models 
similar to the equilibrium ones outlined in Sect.~\ref{subsecScalinglimit} are also 
crucial for the dynamics. In particular, we expect that for `large' values of $\gamma'$ 
the scaling limit condition is not fulfilled perfectly. This is the reason why for the largest
considered $\gamma'$ at small times 
deviations from scaling and even oscillations in the Loschmidt echo are apparent in  
Fig.~\ref{fig:IRLM_com_FRG_DMRG}. 
The amplitude of the latter vanishes for $\gamma'\to 0$ and their frequency is set by the 
bandwidth. We here will not further investigate this non-universal
piece of physics.\cite{comment1} Nonetheless, the curves collapse nicely at large times and 
allow us to access larger rescaled times $T_K t$. In particular, only for sizable 
$\gamma'$ sufficiently large $T_K t$ for analyzing the asymptotic behavior can be reached. 
In general, we find that `small' values of  $\gamma'$ describe well the universal scaling 
curve for 
small $T_K t $, while larger values of  $\gamma'$ leave the scaling limit regime for small 
$T_K t$ but collapse very well for larger values of $T_K t$. It is known from equilibrium that 
sampling the entire scaling function for a finite size lattice model requires the use of 
different values of the impurity strength.~\cite{Enss05} After rescaling by 
$T_K$ the different curves overlap and in combination form the scaling function. We will use this
procedure also for our other impurity models.

\section{Luttinger liquids leads}

\subsection{Single-lead case}

Although the non-equilibrium dynamics of the IRLM studied above involved {\em two non-interacting reservoirs} connected 
through a single-level dot quenching the tunneling between a {\em single interacting} Luttinger liquid lead and 
a quantum dot results in  very similar physics. This motivates why this case is discussed next. 
We consider a reservoir given by Eq.~\eqref{eqXXZlead} ($\alpha={\rm L}$ only) 
whose low-energy physics in the gapless regime ($-1 <\Delta < 1$) falls into the
Luttinger liquid universality class with Luttinger parameter $g^{-1}= 2-\frac{2}{\pi} 
\arccos \Delta$.~\cite{Giamarchi} In the bosonized language,~\cite{Giamarchi} the reservoir can be 
described in terms of a massless compactified boson $\Phi=\Phi_L=\phi^r+\phi^l$ whose right- and left-moving 
components are scattering off the quantum dot. The chiral fields $\phi^r(x)$ and $\phi^l(x)$ live on the interval 
$[0,\infty)$, with the impurity at $x=0$. It is again convenient to unfold the semi-infinite wires to 
define a chiral boson $\phi$ (say, right-moving) on the real line: $\phi(x)=\phi^r(x)$ for $x>0$, and 
$\phi(x)=\phi^l(-x)$ for $x<0$. One then expects that the low energy sector of the Hamiltonian Eq.~\eqref{eqXXZcase} 
is given by 
\begin{equation}
H= \frac{g}{4 \pi}\int_{-\infty}^{\infty}\hspace{-12pt} {\rm d} x 
\, [\partial_x \phi(x)]^2 + \Theta(t) \tilde J'\!    \left( \mathrm{e}^{ - i \phi(0)}  
S_0^{+} + {\rm H.c.} \right)+ \dots,
\label{eqSingleLeadBosonized}
\end{equation}
with $\tilde J' \propto J^\prime$ for small $J^\prime$ and $ \bf{S}_0$ is the auxiliary 
spin of the impurity as 
in the IRLM. The boundary term has dimension $h=\frac{1}{2g} < 1$ and is thus relevant for all values of 
$-1<\Delta < 1$.  Remarkably, this field theoretical Hamiltonian is formally equivalent to that of the two 
lead IRLM Eq.~\eqref{eqIRLMFT}, so all the formulas derived for the IRLM also apply here. In particular, the 
system again flows to a `healed' fixed point at low energy and the corresponding Kondo energy scale reads~\cite{twochannel1}
\begin{equation}
T_K  \propto (\tilde J')^{2 g/( 2 g-1)} \propto (J^\prime)^{2 g/( 2 g-1)}.
\label{eqKondoTempdot}
\end{equation}
In analogy to the IRLM we choose the prefactor to be 4. Further pursuing the analogy, the large time behavior 
of the Loschmidt echo is given by Eq.~\eqref{eqGeneralFormulaLoschmit}  with $\Delta_{\rm BCC}=\frac{1}{8g}$. The entanglement entropy in this case 
is expected to scale exponentially to $\ln(2)$, which signals the onset of the singlet formation between 
reservoir and the dot level. 

However, when `deriving' Eq.~\eqref{eqSingleLeadBosonized}, we dropped the term 
$J^\prime \Delta^\prime \left(\hat n_{1,{\rm L}}- \frac12 \right) \left(\hat n_{0} - 
\frac{1}{2} \right) \sim \partial_x \phi(0) S_0^z$ for $\Delta' \neq 0$ of the microscopic lattice model 
which is {\sl marginal} and could therefore change the exponents in a non-universal way. Within field theory one considers the limit $J^\prime \to 0$. Therefore, since this marginal term scales with $J^\prime$ as well, we do not expect it to modify our results in the field theory limit. In practice however, we also consider sizable $J^\prime$ to cover the full crossover, so one needs to verify numerically the importance of this marginal coupling. Gradually increasing the interaction term $J' \Delta'$ starting at 0, we find numerically that the long-time exponent of the Loschmidt echo 
seems to be independent of this marginal contribution as shown in Fig.~\ref{fig:Singlesite_infDL}. Note however, 
that the time scale for which asymptotic behavior is reached evolves to larger times with increasing $J' \Delta'$.

As an alternative protocol one could consider a setup in which the interaction term 
$ J' \Delta' \left(\hat n_{1,L}- \frac12 \right) \left(\hat n_{0} - \frac{1}{2} \right)$  is present already initially
(only the reservoir-dot hopping is switched on). We expect that in this case the long-time exponent of the 
Loschmidt echo acquires an extra contribution depending on $J' \Delta'$; see Ref.~\onlinecite{SingleLead} for a related 
calculation.

Our DMRG results are presented in Fig.~\ref{fig:Singlesite}. The data for the Loschmidt echo, 
Fig.~\ref{fig:Singlesite} (c) for $\Delta'=\Delta$, collapse rather nicely if rescaled 
by $T_K$ for both positive and negative $\Delta$. Moreover, the long time prediction 
$|G(t)|^2 \sim (T_K t)^{-\frac{1}{2g}}$ --- shown as black dashed lines --- appears to be consistent 
with our data, despite the presence 
of the marginal term  $J^\prime \Delta' \left(\hat n_{1,L}- \frac12 \right) 
\left(\hat n_{0} - \frac{1}{2} \right) $. We find that the data for $\Delta'=0$ collapse 
using the same $T_K$  albeit with 
stronger oscillations; see Fig.~\ref{fig:Singlesite} (a). 
In Fig.~\ref{fig:Singlesite} (b) and (d) we depict the entropy. Scaling is again very 
convincing. The inset of Fig.~\ref{fig:Singlesite} (d) shows that for $t \to \infty$,   
$S(t)$ indeed approaches $\ln(2)$ in an exponential fashion.

\subsection{Dot contact}

\begin{figure}[tb]
 \centering
 \includegraphics[width=0.48\textwidth,clip]{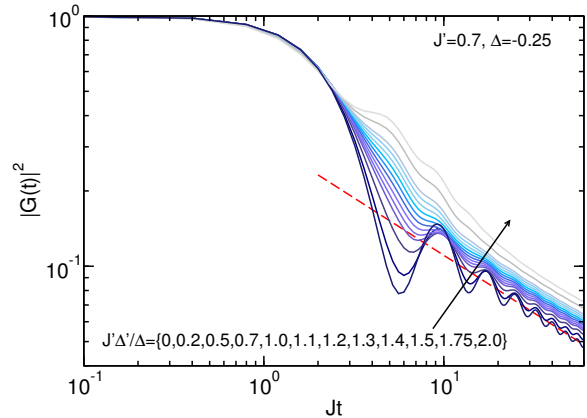}
 \caption{(Color online) DMRG data for $G(t)$ for the dot contact at fixed $J'$.
Gradually turning on the interaction $J' \Delta'$ between the dot and the leads ($L=200$) reveals that the 
marginal term dropped in our field theory analysis of the dot case does not alter the large time exponent found 
(see Fig.~\ref{fig:Singlesite_infDL} for the single-lead setup). The time scales for which asymptotic behavior
can be observed increases with increasing $J'\Delta'$. The field theory prediction 
$|G(t)|^2\sim t^{-\frac{1}{4}\left(1+\frac{1}{g}\right)}$ is included as a dashed red line.}
  \label{fig:Dot_infDL}
\end{figure}
\begin{figure*}[tb] 
 \centering
 
  \includegraphics[width=0.49\textwidth]{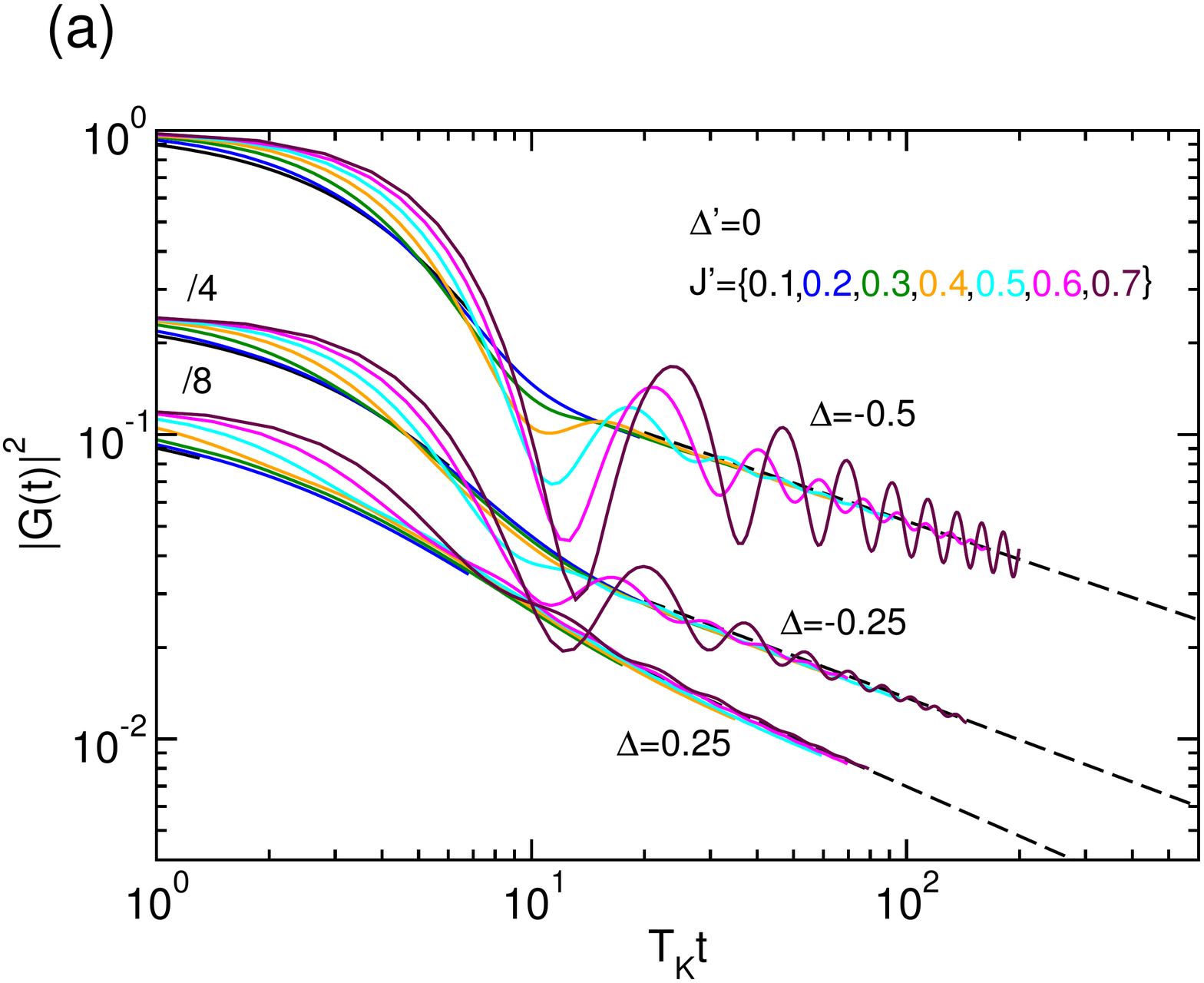}
  \includegraphics[width=0.49\textwidth]{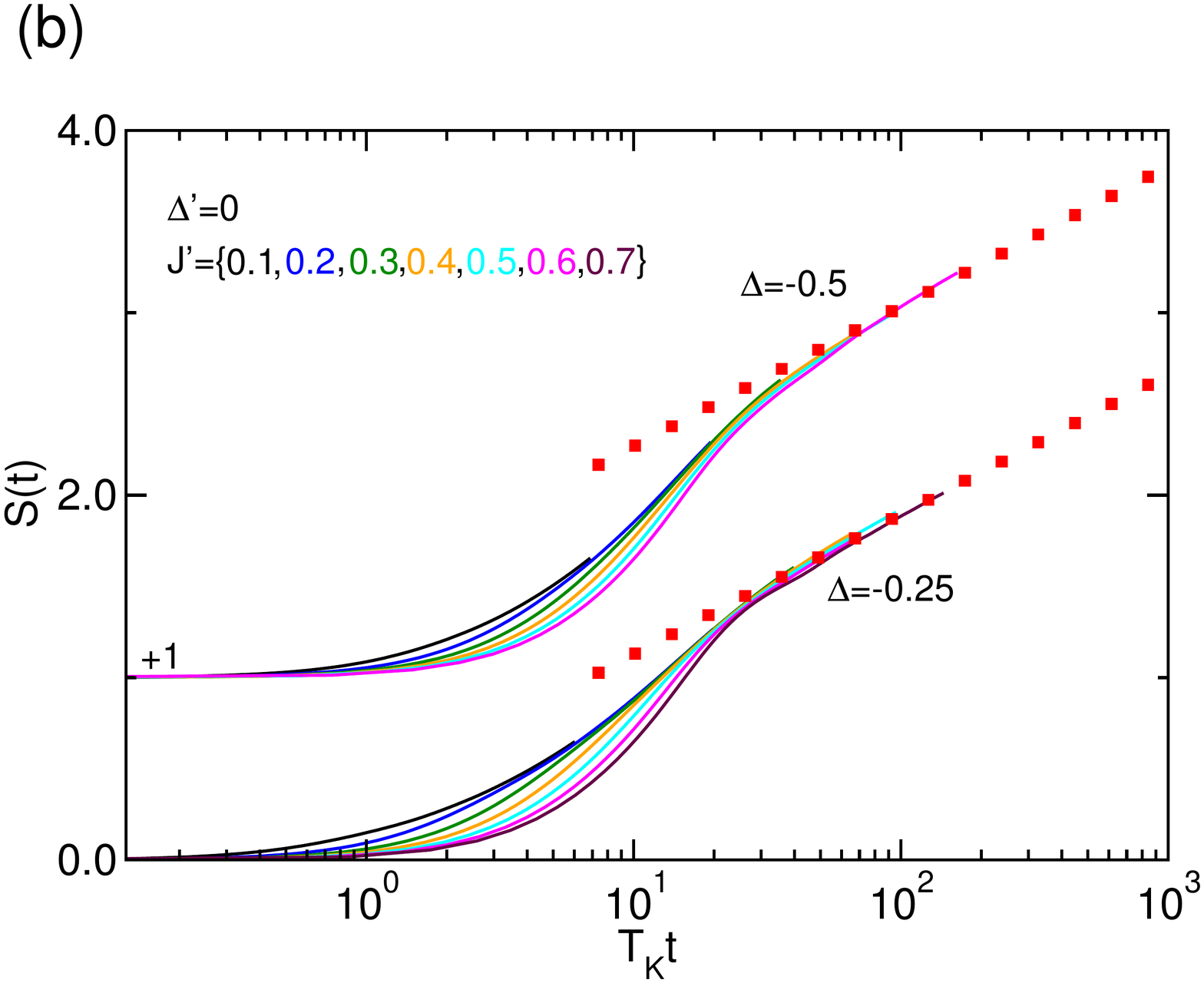}
  \includegraphics[width=0.48\textwidth]{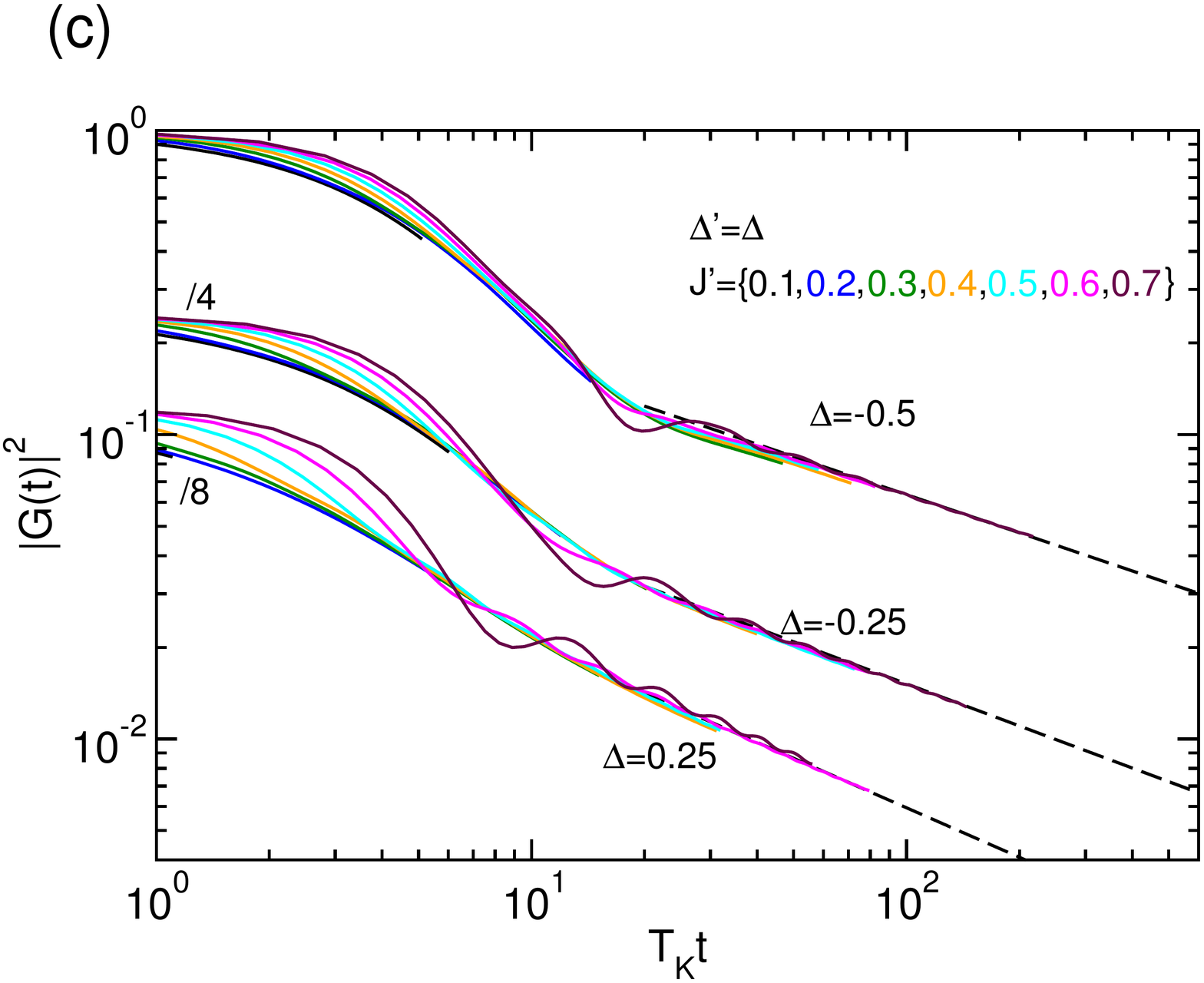}
  \includegraphics[width=0.49\textwidth]{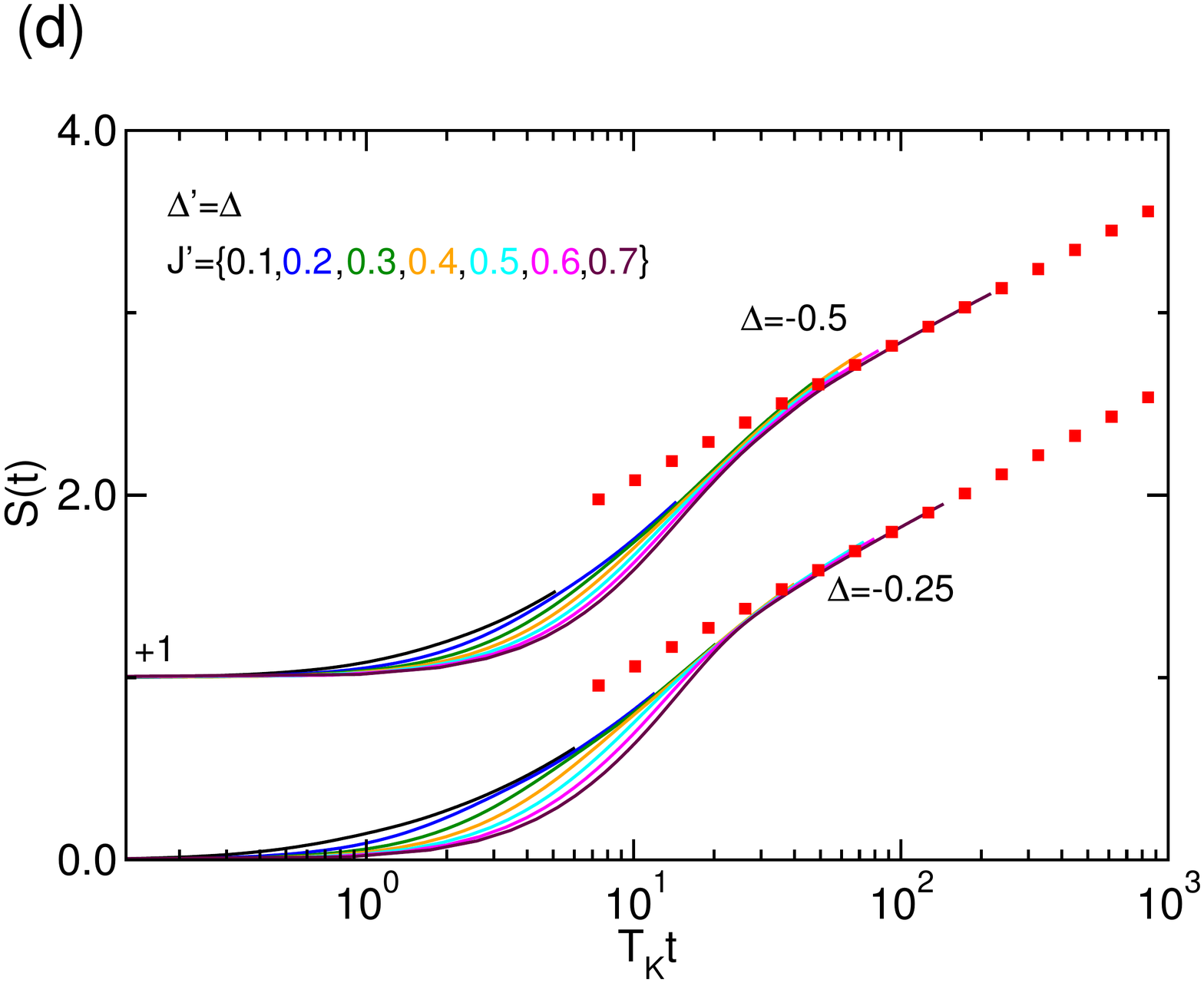}
 \caption{(Color online) Universal scaling of the Loschmidt echo and the entanglement entropy after a quantum quench of 
the tunneling between two Luttinger liquid lead ($L=200$) via a quantum dot. The predicted asymptotic long 
time behavior is indicated as dashed lines. (a), (b) $\Delta'=0$. (c), (d) $\Delta'=\Delta$. (a), (c) 
Loschmidt echo. (b), (d) Entanglement entropy with the large time predictions given by a red dotted line.}
  \label{fig:dot}
\end{figure*}
The low energy limit of the dot contact corresponds to two Luttinger liquid reservoirs connected 
through an effective  spin-$\frac{1}{2}$ impurity $\bf{S}_0$. It can be described in the bosonized 
language starting from the 
single-lead setup discussed above by introducing two bosons (one per channel)
\begin{equation}
H_{\rm{chain},\alpha}= \frac{g}{4 \pi}\int_{-\infty}^{\infty} {\rm d} x \, (\partial_x \phi_\alpha)^2.
\label{eqBulkLL}
\end{equation}
The contact term Eq.~\eqref{eqDotConctact} is then given by
\begin{equation}
\label{eqDotFT}
H_{\rm{link}}(t)= \Theta(t) \tilde J' \left( S_0^{+} \sum_{\alpha=L/R}\mathrm{e}^{ - i \phi_\alpha(0)} 
+ {\rm H.c.} \right)+ \dots,
\end{equation}
with $\tilde J' \propto J^\prime$ for small $J^\prime$. As in the single-lead case, this term has dimension 
$h=\frac{1}{2g}$ ($< 1$ for $g>1/2$ on which we focus) and is thus relevant. The Kondo temperature is given by Eq.~\eqref{eqKondoTempdot}. 
Note that we have again dropped the term $J^\prime \Delta' 
\sum_\alpha\left(\hat n_{1,\alpha}- \frac12 \right) \left(\hat n_{0} - \frac{1}{2} \right) 
\sim\left[\partial_x \phi_L(0)+ \partial_x \phi_R(0)\right] S_0^z$  which is also marginal and could
change the scaling exponent. In analogy to the single lead case we analyze the influence of this marginal 
contribution on the asymptotic exponent of $G(t)$ numerically by increasing $J' \Delta'$ starting at 0. As shown in 
Fig.~\ref{fig:Dot_infDL}, the exponent again appears to be independent of the marginal contribution, although 
the asymptotic time regime is shifted to larger times with increasing $J' \Delta'$.

The low-energy fixed point of the two-lead model is slightly more complex than its single-lead analog. 
It is convenient to introduce the new basis of bosons $\phi_{\alpha={\rm R}/{\rm L}}=\frac{1}{\sqrt{2}} (\phi_\sigma \pm \phi_\rho)$, 
so that the boundary perturbation now reads $\cos \frac{ \phi_\rho(0)}{\sqrt{2}}  
\left(S^+ \mathrm{e}^{ - i \phi_\sigma(0)/\sqrt{2}}  +{\rm H.c.} \right)$. 
At low energy, the value of 
the bosonic field $\phi_\rho(x)$ is pinned down at $x=0$, while $\phi_\sigma$ satisfies a Kondo-like 
boundary condition $\phi_\sigma(0^+)=\phi_\sigma(0^-)+\delta/\sqrt{\pi}$, where 
$\delta^2 =\frac{\pi^2}{8 g} $ (see related discussions in Refs.~\onlinecite{twochannel1,twochannel2,twochannel3}).  
The long-time exponent of the Loschmidt echo Eq.~\eqref{eqGeneralFormulaLoschmit} is thus given by 
$\Delta_{\rm BCC}=\frac{1}{16}+\frac{1}{16 g}$, where the $\frac{1}{16}$ contribution corresponds to changing 
the boundary condition at $x=0$ for $\phi_\rho$ from Neumann to Dirichlet, while the other piece $\frac{1}{16 g}$ 
can be associated with the phase shift $\delta$ for $\phi_\sigma$. For non-interacting leads, the problem reduces 
to the RLM, and one finds $\left| G(t)\right|^2 \sim (T_K t)^{-1/2}$ as expected. When $g=1/2$ (isotropic limit 
$\Delta=1$ of the lattice XXZ chain), the Loschmidt echo decays as $\left| G(t)\right|^2 \sim (T_K t)^{-3/4}$, where 
the value $3/4$ is consistent with the two-channel Kondo orthogonality exponent reported in 
Ref.~\onlinecite{AffleckLudwigFermiEdge}.

We summarize our numerical DMRG results for the dot contact in Fig.~\ref{fig:dot}. Both the Loschmidt echo 
[Figs.~\ref{fig:dot} (a) and (c)] and the entanglement entropy [Figs.~\ref{fig:dot} (b) and (d)] collapse well 
using the Kondo temperature Eq.~\eqref{eqKondoTempdot}, although the Loschmidt echo obtained for 
$\Delta'=0$ shows stronger oscillations again. The entanglement entropy at large times is consistent with 
Eq.~\eqref{eqCFTEntropy}. We find that the large time behavior of the Loschmidt echo is in agreement with 
the  power law prediction Eq.~\eqref{eqGeneralFormulaLoschmit}, with an exponent consistent with the BCFT 
prediction $\Delta_{\rm BCC}=\frac{1}{16} (1+1/g)$ for $\Delta'=0$ as well as $\Delta'=\Delta$.

\subsection{Point Contact}

As a last example, let us consider a point contact Eq.~\eqref{eqPointContact} between two Luttinger 
liquid reservoirs. This case does not involve a dynamical impurity (quantum dot), and the 
bosonized version of the tunneling term reads
\begin{equation}
\label{eqLinkFT}
H_{\rm{link}}(t)=\ \Theta(t)  \tilde J'  \cos  \left[ \phi_{\rm L}(0)-\phi_{\rm R}(0) \right] + \dots,
\end{equation}
where the dots stand for terms being RG irrelevant in equilibrium, and $\tilde J' \propto J^\prime$ for 
small $J^\prime$. We emphasize that Eq.~\eqref{eqLinkFT} corresponds only to the tunneling part 
$\frac{J^\prime}{2} \left(c_{1,{\rm L}}^\dagger c_{1,{\rm R}}+c_{1,{\rm R}}^\dagger c_{1,{\rm L}}\right)$ of 
Eq.~\eqref{eqPointContact}. We have dropped $J^\prime \Delta' \left(\hat n_{1,{\rm L}}- \frac12 \right) 
\left(\hat n_{1,{\rm R}} - \frac{1}{2} \right) \sim \partial_x \phi_{\rm L}(0)  \partial_x \phi_{\rm R}(0)$ as 
it has scaling dimension 2 and is therefore RG irrelevant (in equilibrium).
Forming odd and even combinations, one finds that the even boson decouples while the non-trivial 
part of the dynamics is encoded in a boundary sine-Gordon Hamiltonian for the odd field. The dimension 
of the perturbation is $h=g^{-1}$ such that we shall focus on the attractive regime ($g>1$, $\Delta<0$) 
where it is relevant.~\cite{KaneFisher} To leading order in $\tilde J'$ the RG equation for the amplitude 
of the tunneling term is $d \tilde J'/ d \ell 
= (1-g^{-1}) \tilde J$. For $\tilde J' \ll 1$ the Kondo scale in that case is thus given by 
\begin{equation}
T_K  \propto  (\tilde J')^{g/(g-1)} \propto (J^\prime)^{g/(g-1)}.\label{eq:TK_point}
\end{equation}
In the following, we will set the prefactor to 4 by convention (to match the definition of $T_K$ in the IRLM). 
The large time behavior of the entanglement entropy and the Loschmidt echo in the field theory are given 
by Eqs.~\eqref{eqCFTEntropy} and~\eqref{eqLoschmidtpointcontact}, respectively. This corresponds to having 
$\Delta_{\rm BCC}=\frac{1}{16}$ in Eq.~\eqref{eqGeneralFormulaLoschmit}, which is known to be the scaling dimension 
of the operator changing boundary conditions from Neumann ($\tilde J'=0$) to Dirichlet 
($\tilde J'\to\infty$) in the free boson theory.

\begin{figure}[tb]
 \centering
 \includegraphics[width=0.48\textwidth]{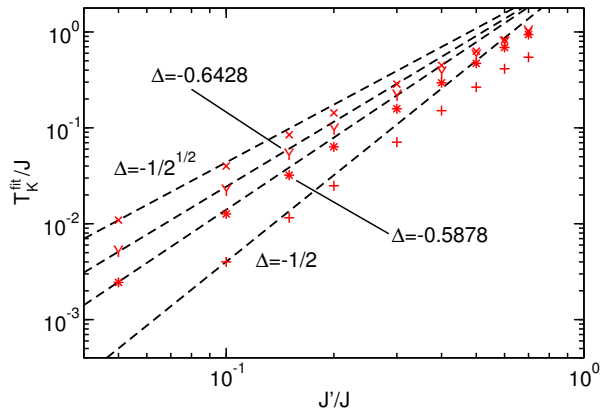}
 \caption{(Color online) Dependence of the Kondo temperature on $J'$ for the point contact and different 
$\Delta$. We show the Kondo temperature $T^{\rm fit}_K$ obtained from collapsing our numerical DMRG data 
by hand (red symbols) as well as the small $J'$ field theory prediction 
$T_K^{\rm FT} \sim (J^\prime)^{g/(g-1)}$ (black dashed lines) on a log-log scale. 
The lines are shifted such that $T^{\rm fit}_K$ and $T_K^{\rm FT}$ coincide at
the smallest $J'$ available. The power-law behavior of $T^{\rm fit}_K$ found for small $J'$ is consistent with the 
prediction from field theory.}
  \label{fig:TKmicro_point}
\end{figure}

\begin{figure*}[tbph!]
 \centering
 \includegraphics[width=0.48\textwidth]{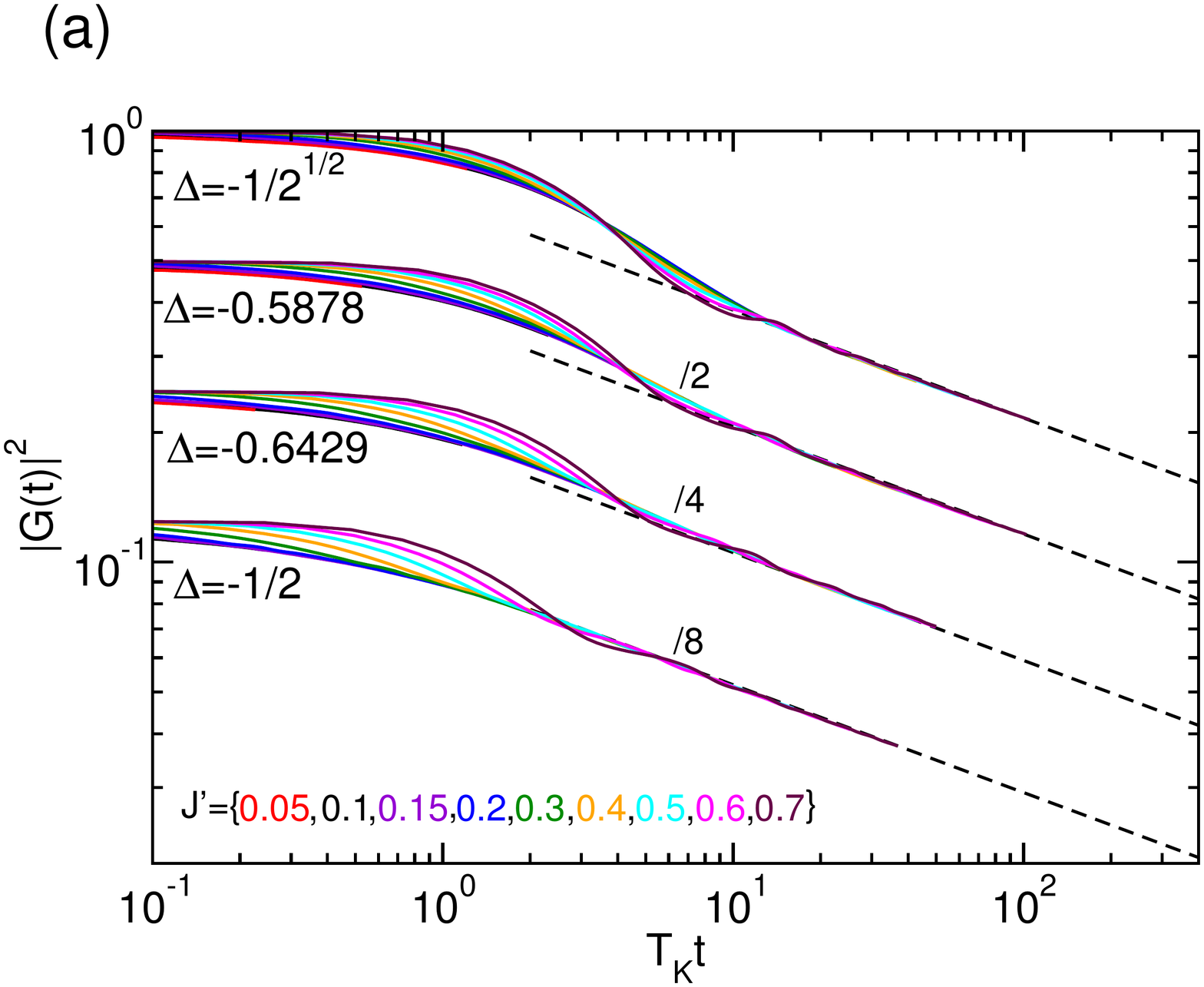}
\hspace{0.1cm}
 \includegraphics[width=0.48\textwidth]{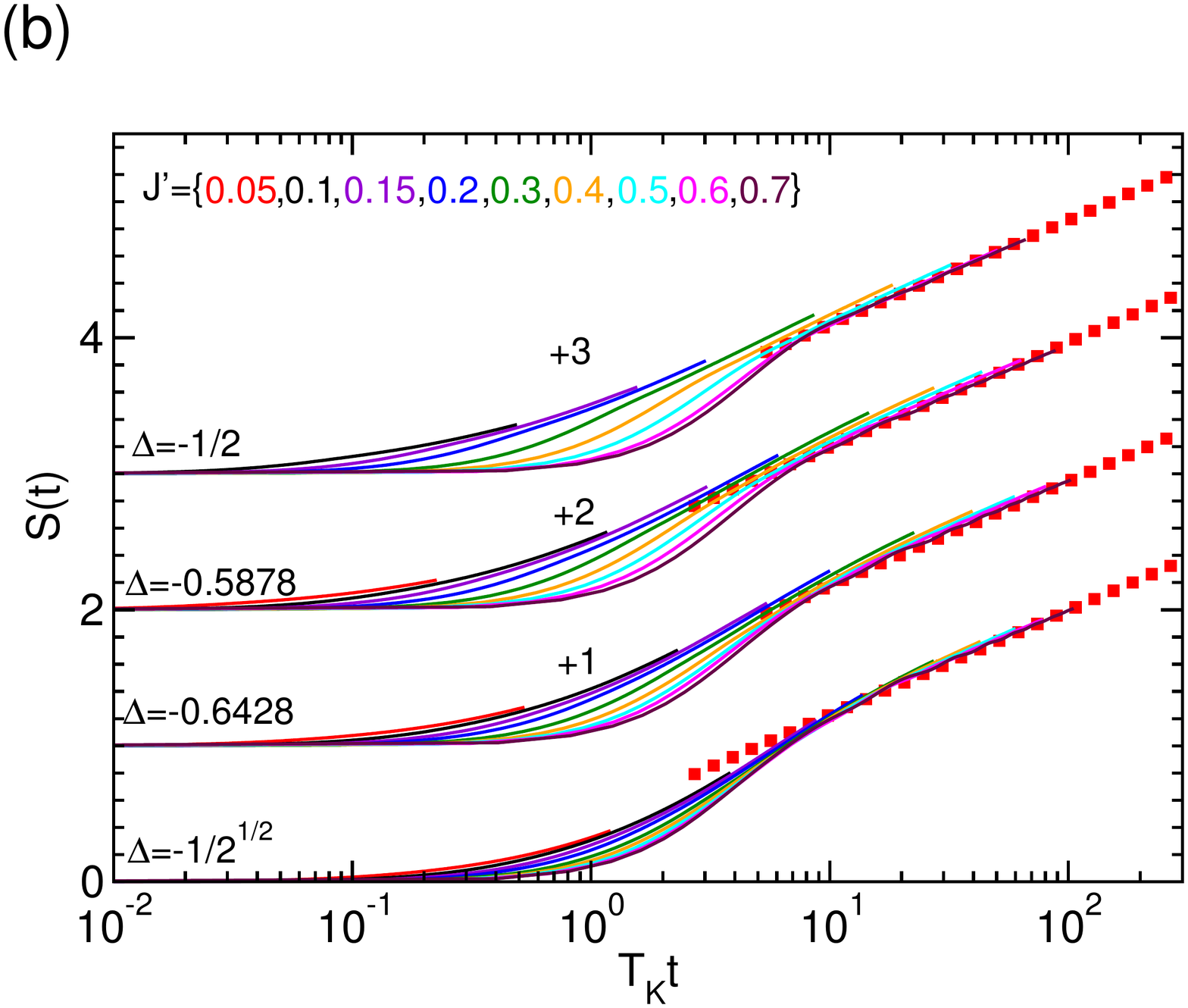}
 \caption{(Color online) Universal scaling of the Loschmidt echo and the entanglement entropy after a quantum 
 quench of the tunneling between two Luttinger liquids connected through a 
 point contact (weak junction). (a) Time evolution of the Loschmidt echo scaled by hand. 
 This  determines the Kondo temperature $T_K$ used in the right plot. The asymptotic long 
 time predictions $|G(t)|^2\sim (T_K t)^{-1/4}$ is shown as a black dashed line. (b)  Universal 
 collapse of the entanglement entropy using the $T_K$ obtained by a collapse of $G(t)$. 
The large  time CFT predictions is given by red dotted lines. }
  \label{fig:pointcontact}
\end{figure*}

To sample the entire scaling function for a lattice with $L=200$ sites we cannot restrict our
considerations to $J' \ll 1$. It is known that for $J' \to 1$ the dual picture of a weak 
impurity, instead of a weak link, is the appropriate one.~\cite{KaneFisher} The dependence 
of $T_K$ on the impurity strength is modified \cite{KaneFisher} and Eq.~\eqref{eq:TK_point} does no 
longer apply. We thus proceed as follows. We consider different $J'$ and scale the DMRG 
post-quench time-evolution data obtained for the Loschmidt echo by hand until they 
collapse; see Fig.~\ref{fig:pointcontact} (a). This gives us an estimate of $T_K$ which
we then also use to rescale $S(t)$.  Approaching small $J'$ we expect to 
find the scaling Eq.~\eqref{eq:TK_point} of $T_K$. The corresponding analysis is shown 
in Fig.~\ref{fig:TKmicro_point}. The field theory prediction works well for small 
$J'$ --- compare the symbols to the dashed lines representing the power law Eq.~\eqref{eq:TK_point} --- 
while rather large deviations can be found for $J' \gtrsim 0.5$ (note the y-axis log-scale). 
This shows that in contrast to the above dot junction cases, in which it was possible to take 
the small hopping field theory expression for $T_K$ (for a detailed analysis indicating 
this for the (I)RLM, see Fig.~\ref{fig:Tkmicro}), 
determining the Kondo temperature by hand is vital for the point contact.

In Fig.~\ref{fig:pointcontact} (a) we show the collapsed DMRG data for $G(t)$ and different
$\Delta<0$. The collapse works particularly well for larger $|\Delta|$ and due to finite size effects
deteriorates for $|\Delta| \to 0$. Remind that we cannot further increase $|\Delta|$ as 
the leading irrelevant {\em bulk terms} will increase and spoil scaling. 
Independent of $\Delta$ the asymptotic behavior is consistent with 
$|G(t)|^2\sim (T_K t)^{-1/4}$ (dashed lines) providing indications for the field theory prediction
$\Delta_{\rm BCC}=\frac{1}{16}$.
We remark that the {\em static} orthogonality exponent $\Delta_{\rm BCC}$ for the same lattice model 
was  studied in~Refs.~\onlinecite{OrthoQinI,OrthoQinII,OrthoVolker}. In the latter work 
the authors employed a very stringent numerical analysis based on a logarithmic derivative. 
In this context the precise numerical evaluation of $\Delta_{\rm BCC}$ remains an open question
due to the difficulty to reach very large system sizes (see also Ref.~\onlinecite{Kennes14a}). 
In our work taking the logarithmic derivative to determine 
$\Delta_{\rm BCC}$ is impractical due to small oscillations prevailing even at large times. 
Therefore, in the light of our analysis, we can only state that the large time exponent 
appears to be consistent with the prediction $\Delta_{\rm BCC}=\frac{1}{16}$. We
certainly cannot rule out small corrections to this exponent, let alone show or disproof 
their existence. Analogous to the problem of reaching very large system sizes (as encountered 
in Ref.~\onlinecite{OrthoVolker}), we would need to analyze much larger times, which is impractical.

Our rescaled DMRG results of $S(t)$ [using the $T_K$ extracted from the by hand scaling 
of $G(t)$] for the point contact are shown in  Fig.~\ref{fig:pointcontact} (b). Again,
the collapse is reasonable. At large times, the data follow the field theory prediction
Eq.~\eqref{eqCFTEntropy}.

\section{Conclusion}
In this paper, we studied the non-equilibrium dynamics of the Loschmidt echo and of the entanglement 
entropy resulting out of abruptly coupling two reservoirs which are  
non-interacting Fermi liquids or interacting Luttinger liquids. 
The coupling is either realized directly by a weak link between the two systems (point contact), 
or indirectly by an additional single site dot in between them (dot contact). 
In addition, we considered the dynamics when coupling a single Luttinger liquid
lead to a single site dot (single-lead case). Microscopic lattice models were used to describe
these setups.
The observables were accessed using both DMRG and FRG. We checked 
numerically the scaling expected from field theory and investigated whether the large time 
behavior can be successfully captured by (boundary) conformal field theory. Simultaneously 
fulfilling the conditions $T_K L/v_F \gg 1$, $T_K / B \ll 1$, and $B^{-1} \ll t \ll L/v_F$ 
under which field theory is expected to describe the physics of lattice models at fixed $L$ 
sets bounds on the strength of the sub-system hoppings entering in $T_K$ as well as on the time $t$. 
Furthermore, the times reachable are bounded from above by the methods used.
Taking this into account, we gathered evidence that for a microscopic 
realization of impurity systems with Fermi or Luttinger liquid reservoirs, 
the dynamics is universal and described by field theory.  It would be interesting to generalize 
our work to finite temperatures, where the work distribution satisfies an out-of-equilibrium 
fluctuation-dissipation relation,~\cite{Crooks} which might bear interesting consequences. The finite temperature generalization of the Loschmidt echo would then be $\langle \mathrm{e}^{ i H_0 t} \mathrm{e}^{ -i H t}  \rangle_0$, where $\langle \dots \rangle_0$ refers to thermal average with the density matrix $\rho_0 = \mathrm{e}^{-H_0/T}/Z_0$, corresponding to the system being held at temperature $T$ before the quench. Because of this new energy scale, we expect both the Loschmidt echo and the entanglement entropy to have a more complicated scaling behavior as functions of $T_K t$ and $T t$. For instance, at small temperature $T/T_K \ll 1$, one can argue from BCFT that the Loschmidt echo for $T_K t \gg 1$ should scale as $\left[ \pi T/\sinh(\pi T t) \right]^{4 \Delta_{\rm BCC}}$, with $T_K$ playing the role of a ultraviolet cutoff. In the large time regime $t \gg T^{-1} \gg (T_K)^{-1}$, this implies an exponential decay of the Loschmidt echo, contrasting with the power-law behavior for $T^{-1} \gg t \gg (T_K)^{-1}$. We leave the detailed analysis of finite temperature quenches for future work.

In addition, we expect that the dynamics resulting out of local quenches
for spinfull impurity systems (spin Kondo effect) should lead to a richer crossover physics. In particular, the Anderson orthogonality exponent dominating the long time behavior of the Loschmidt echo will have contributions coming from both spin channels. Other interesting generalizations of our work include the study of quantum quenches in impurity systems governed by more complicated RG flows, with for instance intermediate fixed points.


\
\

{\it Acknowledgments.} We are grateful to the MPIPKS Dresden for hosting the workshop  
`Quantum Many Body Systems out of Equilibrium' where this work was initiated. 
This work was supported by the Quantum Materials program of LBNL (RV) and the Forschergruppe 723 
of the DFG (DMK and VM). DMK thanks the University of California, Berkeley for hospitality during his visit in summer 2013.
RV also wishes to thank H. Saleur and J.E. Moore for discussions, and the University of Southern California for hospitality 
and support through the US Department of Energy (grant number DE-FG03-01ER45908). 

\providecommand{\newblock}{}

\end{document}